\documentclass[]{aa}

\usepackage{etoolbox}
\makeatletter

\usepackage{natbib}
\bibpunct{(}{)}{;}{a}{}{,} 
\usepackage{caption}
\usepackage{subcaption}
\usepackage{setspace}
\usepackage{graphicx,color}
\usepackage{epstopdf}
\usepackage{setspace}
\usepackage{makeidx} 
\usepackage{lmodern}
\usepackage{packages} 
\usepackage{float}
\usepackage{bm}
\usepackage{amsmath}                  
\usepackage{graphicx}
\usepackage{amssymb}
\usepackage{color}
\usepackage{units}
\usepackage{esint}
\usepackage{braket}
\usepackage{graphics}
\usepackage[dvipsnames]{xcolor}

\newcommand{\rd}{\color{red}}

\def\note #1]{{\bf\rd #1]}}
\newcommand{\eqn}{Eq.}
\newcommand{\eqns}{Eqs}

\def\dd{{\rm d}}

\def \aMLT { \alpha_{\rm{MLT} } }
\def \Yin {Y_0 }
\def \Zin { Z_0 }
\def \YMG { \mathcal{Y} }

\def \Rcz {R_{\rm{cz}} }

\def \Rs{ R_{\odot} }

\def \Teff{ T_{\odot}^{ \rm{eff} } }
\def \ZX{ \left(Z/X \right)_{\rm{surf}} }

\def\be{\begin{equation}}
\def\ee{\end{equation}}

\def\csq{c_{\text{s}}^2}

\begin{document}

\title{Searching for dark energy with the Sun}
\author{ Ippocratis D. Saltas\inst{\ref{praha}}
\and  J\o rgen Christensen-Dalsgaard\inst{\ref{aarhus}}
}
\institute{
CEICO, Institute of Physics of the Czech Academy of Sciences, Na Slovance 2, 182 21 Praha 8, Czechia\label{praha} \and
Stellar Astrophysics Centre, Department of Physics and Astronomy,
Ny Munkegade 120, Aarhus University, 8000 Aarhus C, Denmark\label{aarhus}}

\abstract{General extensions of General Relativity (GR) based on bona fide degrees of freedom predict a fifth force which operates within massive objects, opening up an exciting opportunity to perform precision tests of gravity at stellar scales. Here, focusing on general scalar-tensor theories for dark energy, we utilize the Sun as our laboratory and search for imprints of the fifth-force effect on the solar equilibrium structure. With analytic results and numerical simulations, we explain how the different solar regions offer powerful ways to test gravity. Accounting for the delicate interplay between fifth force and solar microphysics such as opacity, diffusion, equation of state and metallicity, we demonstrate that the fifth force still leaves a sharp signature on the solar sound speed, in a region where simple estimates of input physics uncertainties become negligible. For general scalar-field extensions of GR, known as (U-)DHOST, based solely on the observational helioseismic errors, our analysis at the equilibrium level allows to place an approximate constraint on the fifth-force coupling strength of $-10^{-3} \lesssim \YMG \lesssim  5\cdot10^{-4}$ at $2\sigma$. This result improves previous stellar constraints by $\sim 3$ orders of magnitude, and should be confirmed and improved by future helioseismic inversions in modified gravity combined with an elaborate accounting of theoretical uncertainties. Our analysis can be applied to a wide set of theories beyond GR, and also paves the way for helioseismic analyses in this context. In this regard, we discuss how the solar radiative and convective zone can be employed as promising laboratories to test generic theories of gravity. } 
\vspace{0.5cm}

\maketitle

\section{Introduction}

The need to explain the acceleration of the Universe, and the theoretical shortcomings of the cosmological constant paradigm, have led cosmologists to question the validity of General Relativity (GR) at cosmological scales. The simplest, and widely explored scenario, has been the extension of GR through a new dynamical scalar degree of freedom, which propagates a long-range gravitational force operating on cosmological scales, while it is typically suppressed at local scales to evade the stringent local gravity tests. These are known as scalar-tensor theories and go back to the pioneering work of Brans and Dicke.  Their theoretical structure and phenomenology has been at the forefront of research in astrophysics and cosmology within the last decade. Past years saw the remarkable construction of general classes of scalar-tensor theories known as Beyond Horndeski  \citep{Zumalacarregui:2013pma,horndeski} and DHOST theories \citep{BenAchour:2016fzp,Langlois:2017mxy} respectively. Very recently, these have been extended to the so--called U-DHOST \citep{DeFelice:2022xvq}. 

On the relativistic level, general families of scalar-tensor theories impact on the formation and dynamics of large-scale structures in the Universe \citep[see e.g.][]{Amendola:2017orw, Traykova:2019oyx,Peirone:2019yjs, Hiramatsu:2022fgn}, or the structure of relativistic compact objects \citep[see e.g.][]{Babichev:2016jom,Bakopoulos:2022csr,Baake:2021jzv}. Furthermore, the general class of DHOST theories has been shown to leave a characteristic imprint within Newtonian massive objects, opening up an opportunity to search for dark-energy imprints at stellar scales. It is to be noted that a similar signature within Newtonian stars is expected in a broader context within geometrical extensions of GR which resemble the phenomenology of scalar-tensor theories \citep[see e.g.][and references therein]{Olmo:2019flu}.

As we will explain in more detail below, this local fifth-force effect has been tested in a variety of environments; from white or brown dwarf stars to neutron stars and galaxy clusters. The first work which used the Sun to test this class of theories was presented in \cite{Saltas:2019ius}, who argued that helioseismic observations have the potential to improve on previous stellar physics constraints by two orders of magnitude. In this work, we make an important step further, and study the solar equilibrium structure in the presence of the fifth force, accounting for the delicate solar microphysics, towards accurate and precise helioseismic analyses. We will show how the fifth force leaves a strong signal in the solar interior, detectable with helioseismic inferences of solar interior profiles. 

\begin{figure}
\begin{center}
\includegraphics[width=1 \linewidth]{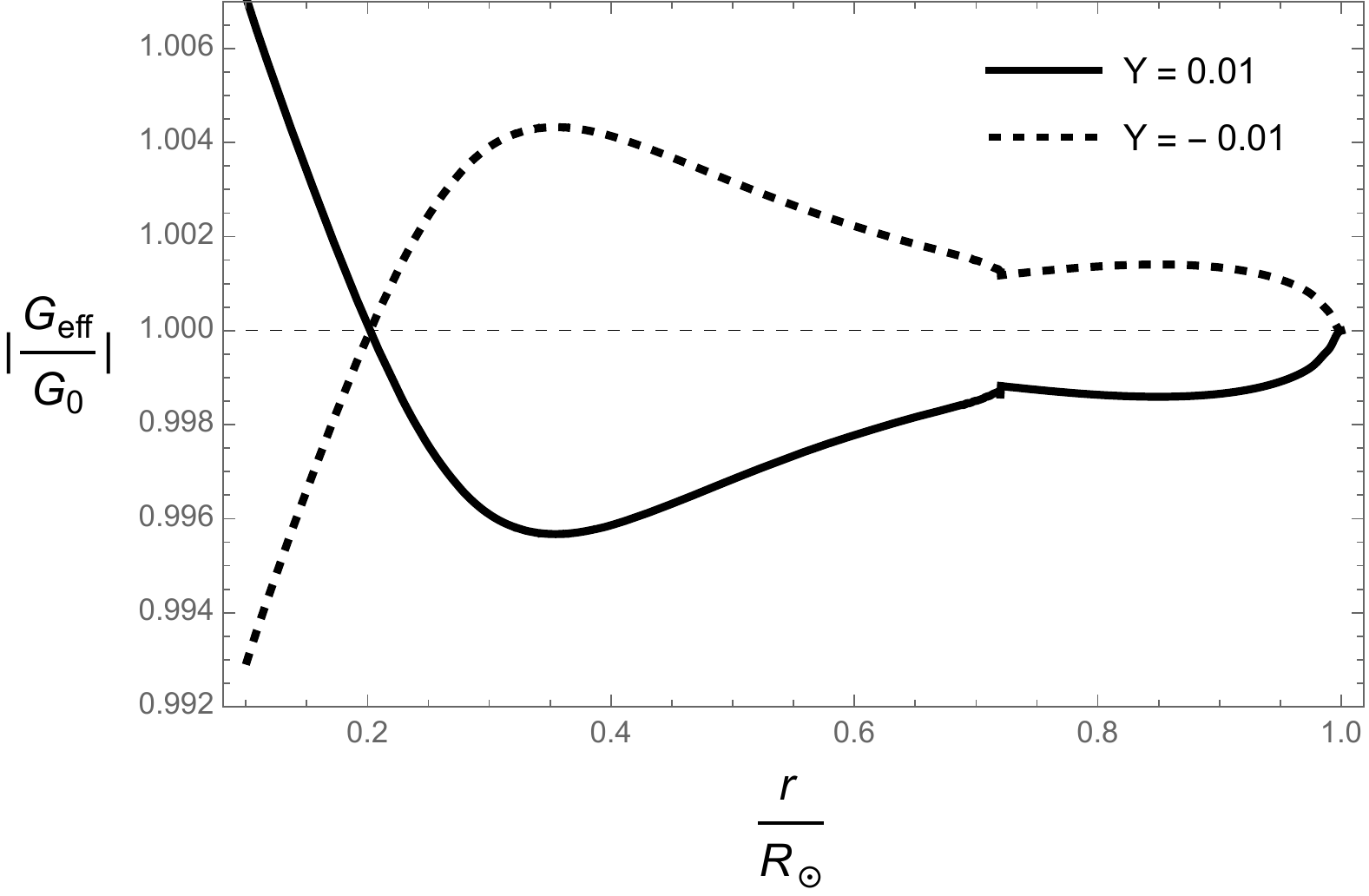}
	\caption{\label{fig:delta_Geff}Effective Newton's coupling induced by the fifth force as a function of radius according to \eqn~(\ref{eq:dP_dr}). Gravity weakens (strengthens) for $\mathcal{Y} > 0$ ($\mathcal{Y} < 0$) for most of the solar interior (see \eqn~(\ref{eq:Geff})), except for $r \lesssim 0.2$, owing to the effects of sperical geometry (see also \eqn~(\ref{eq:d2mass}) below).
The peak of the fifth-force effect around $r \simeq 0.25 R_{\odot}$ in the radiative zone will be key for the phenomenology of solar interior profiles at hydrostatic equilibrium. As expected, the fifth-force effect vanishes at the solar surface . The curves are evaluated according to \eqn~(\ref{eq:Geff}) using the density profile of the reference model GS98 of Table \ref{table:models}, computed without fifth force. }
\end{center}
\end{figure}

Within the general scalar-tensor extensions of GR aiming to explain dark energy, known as DHOST theories, the weak-field limit of the theory predicts that the hydrostatic equilibrium of massive objects is modified through a new fifth-force term \citep{Kobayashi:2014ida,Crisostomi:2017lbg,Dima:2017pwp} as
\begin{align}
\frac{\dd P(r)}{\dd r} = - \rho(r) \frac{G_{\rm{eff}}(r) m(r)}{r^2}, \label{eq:dP_dr}
\end{align}
where $m(r)$ is the mass enclosed within a radius $r$ from the centre and $\rho(r), P(r)$ the density and pressure respectively. Within Newtonian gravity one obviously has $G_{\rm{eff}} = G_{0}$, with $G_{0}$ the bare Newton's constant measured in the solar system. As soon as the fifth force operates, the effective Newton's coupling on the level of hydrostatic equilibrium is promoted to a function of the radius as 
\begin{align}
G_{\rm{eff}} \equiv \left(1 + \frac{\mathcal{Y}}{4}r^2 \frac{m''(r)}{m(r)} \right) G_{0}, \label{eq:Geff}
\end{align}
with $' \equiv \dd/\dd r$, and $\mathcal{Y}$ the dimensionless fifth-force coupling. The coupling $\mathcal{Y}$ relates parametrically to those parameters governing the theory's dynamics around a cosmological background. Since $m''(r) < 0$ within most of the star, relation (\ref{eq:Geff}) suggests that gravity tends to weaken (enhance) for $Y>0$ ($Y<0$). A plot of \eqn~(\ref{eq:Geff}) for indicative values of $\YMG$ is shown in Figure~\ref{fig:delta_Geff}. The effect of the fifth force acquires its maximum value around $r \sim 0.3 R_{\odot}$, and it switches off at the solar surface where $m''(r) = 0$. 

It is interesting to note that a similar modification of gravity's strength is predicted in different theoretical setups (\cite{Olmo:2019flu, Wojnar:2022txk}). Indeed, modifications to the Newtonian law from some extended model of gravity is expected to typically involve gradients of density. Therefore, the results of our analysis can be appropriately adopted to apply to more general theory setups. 

Earlier works constraining the fifth-force coupling $\mathcal{Y}$ have been carried out in the context of white dwarfs \citep{Jain:2015edg, Saltas:2018mxc}, red/brown dwarfs \citep{Sakstein:2015zoa, Sakstein:2015aac,Kozak:2022hdy}, main-sequence stars \citep{Sakstein:2016lyj}, relativistic compact objects \citep{Babichev:2016jom}, galaxy clusters \citep{Sakstein:2016ggl, Pizzuti:2019wte, Pizzuti:2021brr, Laudato:2021mnm, Haridasu:2021hzq}, Hulse-Taylor pulsar \citep{Dima:2017pwp}, gravitational waves \citep{Creminelli:2018xsv}, solar system \citep{Crisostomi:2019yfo}\footnote{See also \cite{Koyama:2015oma, CANTATA:2021ktz, Saltas:2021eff}.}. All constraints from stellar scales have so far imposed $\YMG < \mathcal{O}(0.1)$ at $2\sigma$. 

A first exposition on the potential of the Sun and helioseismology to constrain this general class of dark-energy theories has been presented in \cite{Saltas:2019ius}. In that work, a polytropic EoS was used to predict the solar pulsation frequencies within the Cowling approximation, showing that helioseismology has the potential to tightly constraint the fifth-force coupling at the $10^{-3}$ level. 
Our present goal is to provide an accurate, quantitative description of how the fifth force affects the solar interior towards precision constraints on the fifth force with helioseismology. 
The work splits into two parts. In the first part (Section~\ref{sec:theory}), we study the effect of the fifth force on solar observables by means of intuitive arguments and analytic relations. In the second part (Sections~\ref{sec:calibration}, \ref{sec:systematics}), we numerically investigate the evolution of the present Sun with numerical simulations, and analyse the interplay between fifth force and solar input physics. The focus is on the effect on the solar interior profile, most notably the sound-speed, which is the key to the precise science of helioseismology. We will show that the fifth-force effect gives a {\it characteristic effect} deep in the interior zone, at a point where uncertainties from opacity or diffusion become negligible, thus making it distinct within helioseismic analyses. In Section~\ref{sec:constraint} we derive an order of magnitude constraint for the fifth-force coupling based on its effect on the solar sound speed, and in Section~\ref{sec:summary} we draw our conclusions.

\section{Fifth force in the solar interior: Theoretical considerations} \label{sec:theory}

The dynamics of stellar evolution are modelled by the equations describing hydrostatic equilibrium, energy conservation and transport, coupled to the time evolution of element abundances in the star. The system of equations closes under an assumption on the equation of state and other microphysics, in particular, the opacity, diffusion, and the treatment of convective energy transport. Although assumptions on microphysics do not affect gravity directly, they may implicitly interfere with it through their impact on the stellar interior profiles such as density and pressure. Let us recap the basic equations and discuss the way they are affected by the presence of a fifth force. 

Overall, a weakening (strengthening) of the gravitational constant $G$ in the solar interior will lead to a change in pressure, as can be easily seen using the equation of hydrostatic equilibrium to estimate $P_{\text{central}} \sim G M_{\odot}^2/R_{\odot}^4$. This will be followed by a change in the temperature of similar magnitude, which can be estimated assuming a perfect gas EoS, $P \propto \rho T$. Obviously, a weakening (strengthening) of gravity will tend to make the star less (more) dense. In order to retain hydrostatic equilibrium at the same radius, the star will inevitable adjust its structure, which affects the energy generation from nuclear reactions, as we will discuss later.

The energy equation relates the spatial gradient of luminosity to the energy generation, and will be not explicitly affected by the change in gravity. The same holds for the equation describing the change of element abundances with time. Both equations, however, will be affected implicitly through the change in the core's temperature, density and initial element abundance. To see this, let us write explicitly the equation of energy transport as
\be
\frac{\dd T}{\dd r} = \nabla \frac{T}{P} \frac{\dd P}{\dd r}, \label{eq:dT_dr}
\ee
with the gradient $\nabla \equiv \dd\log T/\dd\log P$. For radiative transport of energy, $\nabla = \nabla_{\rm{rad}} \propto \kappa(r) \, P(r) L(r)/(m(r) T^4)$, where $\kappa$ is the opacity and $L$ the luminosity. In view of the modified hydrostatic equilibirum (\eqn~(\ref{eq:dP_dr})), \eqn~(\ref{eq:dT_dr}) implies that
\be
\left| \frac{\dd T}{\dd r} \right|_{\YMG  {}^{>}_{<} 0}    {}^{<}_{>}  \left| \frac{\dd T}{\dd r}\right|_{\YMG = 0},
\ee
i.e. a weakening of gravity ($\YMG>0, G_{\rm{eff}}/G_{0} < 1 $) will tend to decrease the temperature gradient, and the opposite is true for stronger gravity ($\YMG<0, G_{\rm{eff}}/G_{0} > 1 $). Therefore, a weaker (stronger) gravity will act as to make the energy transfer less (more) efficient. This argument serves only as a qualitative way towards gathering first intuition, since the fifth force affects hydrostatic equilibrium in a non-trivial, radius-dependent fashion. 

The equation describing the evolution of element abundances with time will not receive explicit corrections, but will be implicitly affected through the change in the energy generation rate, which for the pp chain is
$
\epsilon_{\text{pp}} \propto X_{\text{H}}^2 \rho T^{6}.
$
The change in the central temperature, density and hydrogen abundance will induce a change in the energy generation. The change in the initial hydrogen abundance ($X_{\text{H}}$) follows from the need to maintain the same luminosity at the present solar age, as we will discuss later on. 

\subsection{Density profile}
Let us start by understanding the behaviour of the solar interior profiles from a quantitative point of view, through the derivation of approximate, and whenever possible, analytical solutions. An often useful EoS to model a star in a simple manner is the polytropic relation defined through
\be
P = K \rho^{\Gamma_1} \equiv K \rho^{1+ \frac{1}{n}}, \label{eq:Polytrope}
\ee
where $K$ is the adiabatic constant, $\Gamma_1, n$ the (constant) adiabatic and polytropic index respectively, and $P, \rho$ the pressure and density profiles. The solar radiative zone is crudely described by $n=3$ and here the polytropic equation needs to be solved numerically. 
In the convective zone, \eqn~(\ref{eq:Polytrope}) is approximately satisfied, with $\Gamma_1$ being the adiabatic compressibility, 
$\Gamma_1 = (\partial \log p/\partial \log \rho)_{\rm ad}$; in the bulk of the Sun matter is approximately in the form of a fully ionized ideal gas, with $\Gamma_1 \simeq 5/3$, corresponding to $n = 1.5$. In this case an approximate analytic solution can be found, as discussed below.

Polytropic solutions for $n = 3$ in a solar context were previously discussed in \cite{Saltas:2019ius}. As was shown there, the effect of the fifth force on the sound-speed profile in the radiative zone grows from the centre until it peaks around $r \simeq 0.3 R_{\odot}$, similar to the scaling exhibited by the effective gravity strength in Figure~\ref{fig:delta_Geff}. A similar behaviour is expected for the fractional change of the density profile under the fifth force, and confirmed by the evolutionary simulations we performed in this work, shown in Figure~\ref{fig:model_differences}. To understand the aforementioned behaviour of the sound speed in the radiative zone quantitatively, we model the radiative interior as an ideal gas, $P \propto \rho k_{\rm B} T$, such that
\be
c^{2}_s(r) \equiv\Gamma_1 \frac{P}{\rho} \simeq \frac{\Gamma_1 k_{\rm B} T}{\mu m_{\rm u}}, \label{eq:cs-ideal}
\ee
where $k_{\rm B}$ is the Boltzmann constant, $\mu$ the mean molecular weight, and $m_{\rm u}$ the atmomic mass unit.
This approximation will be useful later on when we interpret the results of our numerical simulations.  The adiabatic index $\Gamma_1$ intimately relates to the properties of the EoS, and the intuition suggests that the effect of the fifth force on it will be minor, and only indirect through the change in the density and pressure.
Therefore, the main effect on $\delta c^{2}_s/ c^{2}_s$ in the radiative zone will be driven by the change in the temperature and mean molecular weight,\be
\frac{\delta c_{s}^2(r)}{c^{2}_{s} }\simeq \frac{\delta \Gamma_1}{\Gamma_1} + \frac{\delta T}{T} - \frac{\delta \mu}{\mu}. \label{eq:cs-frac-ideal1}
\ee
We discuss this in detail in Section~\ref{sec:systematics}.

Let us now turn our focus on the derivation of analytic solutions for solar profiles in the solar convective zone. The mass enclosed in a spherical shell up to a radius around the base of the convective zone is fairly close to the total solar mass, so we may write for the mass profile in this region
\be
m(r) \simeq M_{\odot} + \mathcal{O}\left(\frac{\delta m}{M_{\odot}} \right), \; \; \; \delta m = M_{\odot} - m(r>\Rcz),
\ee
with the neglected terms sufficiently small, given that the convective zone accounts for $2.5 \%$ of the solar mass. We first examine the solution for $\YMG = 0$. Assuming $m(r) \simeq M_{\odot}$, we  approximate the r.h.s of the equation of hydrostatic equilibrium as
\begin{align}
	\frac{\dd P}{\dd r} & =  K  \Gamma_1 \rho^{\Gamma_1-1} \frac{\dd \rho}{\dd r} = - \frac{G_{0} M_{\odot} }{r^2} \rho \, \left[ 1 - \mathcal{O}\left(\frac{\delta m}{M_{\odot}}\right) \right], \label{eq:hydro_CZ}
\end{align}
with the second equality due to the polytypic EoS.
Assuming $\Gamma_1$ to be constant, as is approximately true except in the near-surface ionization zones of hydrogen and helium, and neglecting the small correction on the r.h.s, we can straightforwardly integrate the above equation to find
\begin{align}
\rho_{(0)}(r) \equiv B \left( \frac{R_{\odot}}{r} - 1 \right)^{\frac{1}{\Gamma_1-1}} \,  B \equiv \left( \frac{\Gamma_1 - 1}{\Gamma_1} \frac{GM_{\odot}}{K \, R_{\odot}} \right)^{\frac{1}{\Gamma_1 - 1}},\label{eq:rho0-solution}
\end{align}
with $B$ a constant with dimension of density. This solution describes approximately the density profile in the solar convective zone. Notice that in principle this solution needs be matched with 
conditions in the radiative zone at the convective boundary. 

We now consider the extension of the previous analytic expression to the case where the fifth force is turned on, $\YMG \neq 0$. Since the fifth-force coupling is small, we will adopt a perturbative approach. In particular, we start from the general hydrostatic \eqn~(\ref{eq:hydro_CZ}), and assume a solution for the density profile in the form
\be
\rho(r; \YMG) \simeq \rho_{(0)}(r) + \epsilon \rho_{(\YMG)}(r), \label{eq:rho_expansion}
\ee
where $\epsilon \equiv \pi \YMG$ is a small parameter $\ll 1$, and $\rho_{(\YMG)}(r)$ the small contribution on the density profile from the fifth force. A similar perturbative solution will hold for the sound speed, which we derive later. To ease the computation, instead of deriving the solution for any $\Gamma_1$, we now restrict ourselves to $\Gamma_1 = 5/3$, which corresponds to the choice of a polytropic index $n = 3/2$. Plugging \eqn~(\ref{eq:rho_expansion}) back into the hydrostatic \eqn~(\ref{eq:hydro_CZ}), the $\epsilon$-dependent part of the equation takes the form
\be
z(1-z)\rho_{(\YMG)}' - \rho_{(\YMG)}  + P(z) = 0, \; \; \; z \equiv \frac{r}{\Rs}, \label{eq:diff-eq-rhoY}
\ee
with $' \equiv d/dz$, and $P(z) \propto 1/2 - 3z + (9/2)z^2 - 2z^3$, modulo an overall normalisation constant. The homogeneous solution is easily found first setting $P(z) = 0$, and then integrating through separation of variables. This yields 
\be \rho_{(\YMG)}^{\rm{homog.}} = C \,\frac{z}{1-z},
\ee
with $C$ a constant of integration. Since the full solution will be given by the sum of the homogeneous and a inhomogeneous solution, to find the latter, we seek a solution of \eqn~(\ref{eq:diff-eq-rhoY}) in the form $\rho_{(\YMG)}^{\rm{inhomog.}} = \rho_{(\YMG)}^{\rm{homog.}} Q(z)$, where $Q$ a function to be determined. Substituting this ansatz in \eqn~(\ref{eq:diff-eq-rhoY}), one arrives at the differential equation $Q' = -P(z)/z^2$, which can be easily integrated to find $Q$, since the explicit form of the polynomial $P(z)$ is known. After choosing the integration constant $C$ in the homogeneous solution so that the solution for \eqn~(\ref{eq:diff-eq-rhoY}) is well-behaved on the solar surface, i.e. no divergence occurs as $z \to 1$, one finally finds
\be
 \epsilon \cdot \rho_{(\YMG)}(r) \simeq c \, \left( -1 - 7 z + 2 z^2 + \frac{6z\log(z)}{z-1} \right), \label{eq:rho1-solution}
\ee 
with the constant $c$ defined as 
$$
c \equiv \YMG \left(\frac{3 \pi}{10} \frac{B^{4/3}G \Rs^2 }{\Gamma_1 K } \right),
$$
and we also re-introduced our expansion parameter $\epsilon \equiv \pi \YMG$, appearing in \eqn~(\ref{eq:rho_expansion}). The solution (\ref{eq:rho1-solution}) is understood as a small correction on top of the $\YMG=0$ solution, $\rho_{(0)}$, and it can be seen that it vanishes on the solar surface, i.e. in the limit of $z \to 1$.
One further notices the different scaling with radius $z \equiv r/\Rs$ compared to the one derived in standard gravity ($ = 3/2$), \eqn~(\ref{eq:rho0-solution}). We now define the fractional change of density as
\be
\frac{\delta \rho}{\rho} \equiv \frac{\rho(\text{standard model}) -  \rho(\text{modified model})}{\rho(\text{standard model})}, \label{eq: deltac}
\ee
with $\rho = \rho(r)$ understood as in \eqn~(\ref{eq:rho_expansion}). The definition (\ref{eq: deltac}) will serve as a definition for the fractional change of interior profiles under any change of physics.
It is straightforward to check that $\delta \rho / \rho$, like  $\rho_{(\YMG)}$, also vanishes on the solar surface. To understand the qualitative behaviour of the fifth-force term on the fractional change of density, we may expand to first order in $z^{-1} \equiv R_{\odot}/r$ around the base of the convective zone which we conventionally choose as $R_{\text{cz}} \simeq 0.71 \Rs$,
\be
\frac{\delta \rho}{\rho}(r \gtrsim R_{\text{cz}}) \simeq -0.57 \, \YMG - 0.74 \, \YMG \left( \frac{ r - \Rcz}{R_{\odot}} \right).
\ee
For the typical value of $\YMG = 0.01$, the above solution implies a very slow change in the convective zone, which is confirmed by our detailed solar evolutionary simulations shown in Figure~\ref{fig:model_differences}.

\subsection{Sound speed}
We now look at  the sound-speed profile defined through 
\be
c^{2}_s(r) \equiv \Gamma_1 \frac{P}{\rho} \approx K \Gamma_1 \rho^{\Gamma_1 -1}, \label{eq:cs-def}
\ee
with the last approximate relation is valid for a polytrope. Using the analytic solutions for the density derived earlier, it is easy to derive an expression for the sound speed in the convective zone and at zero fifth force ($\YMG = 0$) as 
\begin{align}
c_{s (0)}^2(r) \simeq \frac{G_{0}M_{\odot}}{R_{\odot}} \, (\Gamma_1 -1) \, \left( \frac{\Rs}{r} -1 \right), \label{eq:c^2-convective}
\end{align}
reminding that we neglected corrections due to the non-constancy of the mass profile in the solar convective zone. It is also understood that this relation needs be matched with the polytropic one in the radiative zone, which will fix the polytropic constant $K$ -- we omit doing this here, since we are only interested in understanding the qualitative scaling.  The solution (\ref{eq:c^2-convective}) was first derived in \cite{JCD:1991}.

\begin{figure} 
    \centering
       \begin{subfigure}[t]{0.47\textwidth}
        \centering
        \includegraphics[width=\linewidth]{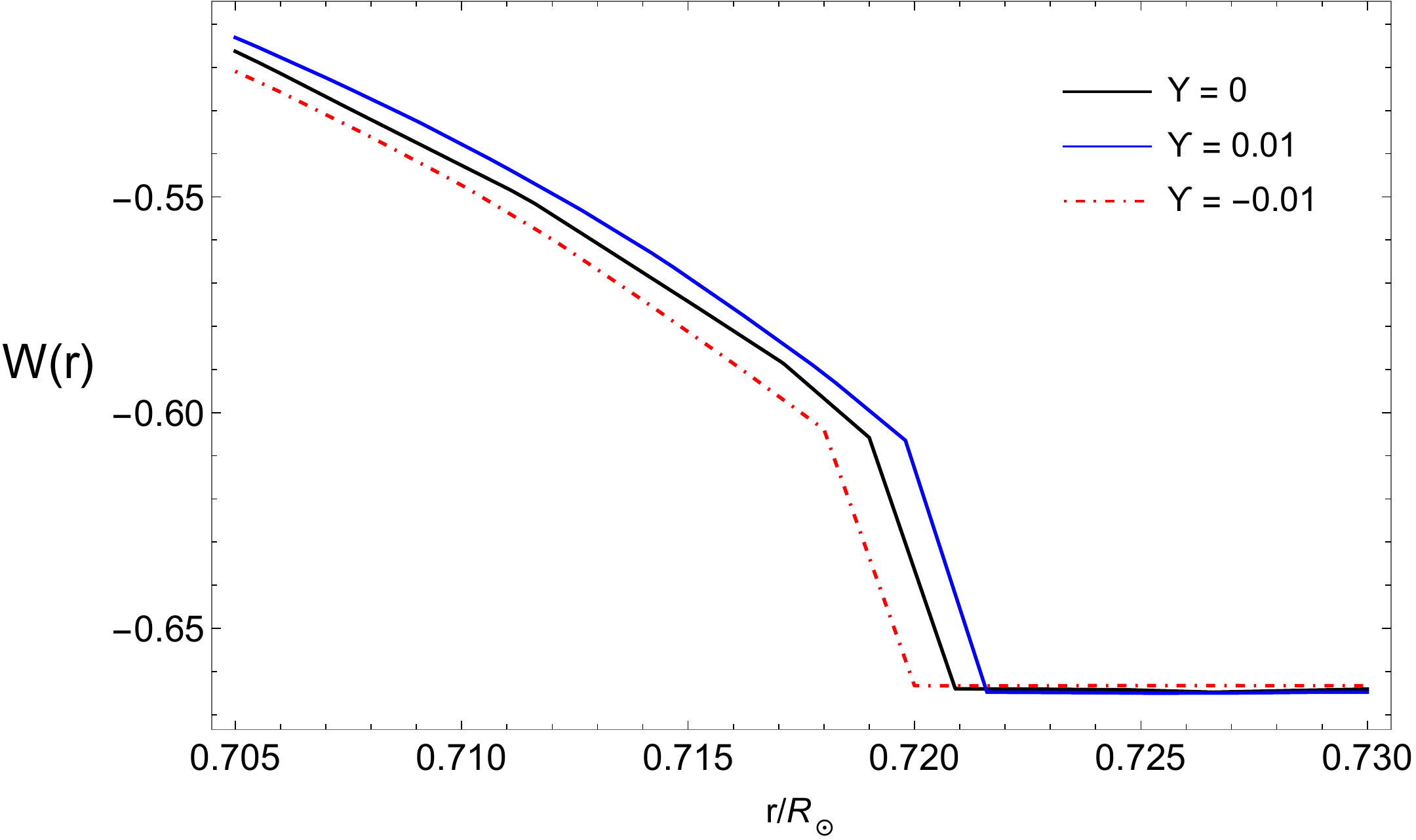} 
    \end{subfigure}
    \begin{subfigure}[t]{0.47\textwidth}
        \centering
        \includegraphics[width=\linewidth]{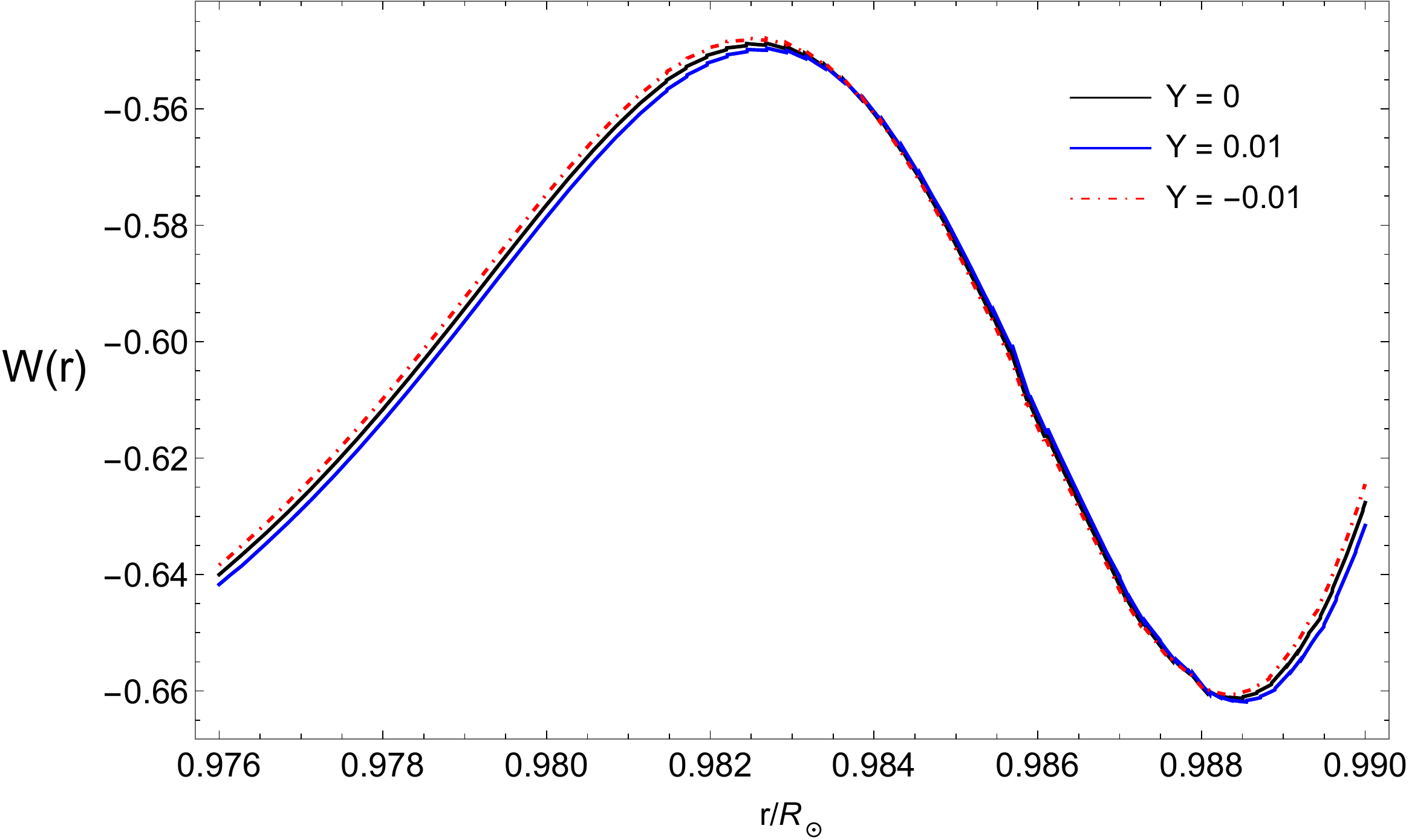} 
    \end{subfigure}
    \hfill
	\caption{\label{fig:W} {\bf Top}: The quantity $W(r)$, defined in \eqn~(\ref{eq:W(r)}), is a measure of the sound-speed gradient and a key diagnostic for the solar structure. Here, it is shown for the following three cases: for standard gravity ($\YMG = 0$), weaker ($\YMG = 0.01$) and stronger ($\YMG = -0.01$) gravity respectively. The plots are based on full numerical simulations of {\it calibrated} models for radius, luminosity and surface metallicity (see Table \ref{table:models} for a summary). The solution of $W(r) = -2/3$ yields an approximate estimate of the position of the radius of the base of the convective zone ($r \simeq \Rcz$), which is shown in the upper panel. The reference model at standard gravity is model GS98 summarised in Table \ref{table:models}. As it can be seen from the upper panel, a weaker (stronger) gravity tends to make the convective zone shallower (deeper). {\bf Bottom}: The peak of the quantity $W(r)$ right below the solar surface due to the ionisation of helium. Although barely visible by eye, the position of the peak is shifted inwards (outwards) for stronger (weaker) gravity.   }
\end{figure}

If we switch on the fifth force, the relevant sound-speed profile is then straightforwardly found to be 
\be
c_{s}^2(r; \YMG)  \simeq K \,  \Gamma_1 \, \left(\rho_{(0)}(r) + \epsilon  \rho_{(\YMG)}(r) \right)^{\Gamma_1 -1}, 
\ee
which applies, under our assumed approximations, in the convection zone. Thus, here the speed of sound will inherit the density's approximate constancy, which is also seen in the full numerical solutions of calibrated solar models in Figure~\ref{fig:model_differences}. A key observation from \eqn~(\ref{eq:c^2-convective}) is that the sound-speed profile is independent of the adiabatic constant $K$, making it rather insensitive to change of input physics such as opacity/diffusion, in the convective zone. Similar to what we did  before, we expand to first order around the base of the convective zone ($\simeq 0.71 R_{\odot}$), assuming again $\Gamma_1 = 5/3$, to find 
\be
\frac{\delta \csq}{\csq} (r \gtrsim R_{\text{cz}})\simeq 
0.1 \YMG - 0.5 \YMG \left( {r- R_\text{cz} \over {\Rs}} \right).
\label{eq:delcsq}
\ee
This explains the mild variation of $\delta \csq(r)/\csq$ for $r \gtrsim R_{\text{cz}}$. The gradient of the sound speed forms a critical diagnostic for solar structure in the convective envelope, and it is usually defined through the quantity $W(r)$  \citep[e.g.,][]{Gough:1984, Dappen_Gough:1986} as
\be
W \equiv \frac{r^2}{G_{0}m(r)} \frac{\dd  \csq}{\dd r} \simeq \frac{r^2}{G_{0}M_{\odot}} \frac{\dd  \csq}{\dd r}, \label{eq:W(r)}
\ee
with the last approximate equality true within the convective zone. In the convective zone, $W \simeq 1 - \Gamma_{1}$, with $\Gamma_1 \simeq 5/3$ and therefore $W \simeq -2/3$. This in turn can be used to infer the thickness of the convective envelope if the sound-speed profile is known through helioseismic observations \citep{JCD:1991}.
The applicability of this procedure in the presence of the fifth force relates to whether $\Gamma_{1}$ receives sizable corrections from the fifth force, such that the value of $W$ would depart significantly from $-2/3$. Our simulations, to be presented below, show that under the fifth force $\delta \Gamma_{1}/ \Gamma_{1} \sim \mathcal{O}(10^{-5})$, and therefore the effect is negligible.  In Figure~\ref{fig:W} we show $W(r)$ for {\it calibrated} solar models within standard and modified gravity. It is seen that a stronger gravity in the convective zone ($\YMG<0$) will act as to shift the bottom of the convective zone to a lower point compared to standard gravity, and the opposite is true for $\YMG>0$.  In other words, a stronger gravity implies a steeper sound-speed profile.

\section{Solar calibration method in the presence of fifth force} \label{sec:calibration}
\subsection{Numerical method}
We use the 1-dimensional MESA stellar evolution code \citep{Paxton2011,Paxton2013,Paxton2015}. The code solves the system of stellar equations discretely at each time step on a 1-dimensional spatial grid, assuming quasi-hydrostatic equilibrium and spherical symmetry. Solutions to the equations are then sought using a variant of the Newton-Raphson method known as the Henyey method \citep{henyey1964}. The equations of stellar structure form a boundary value problem with boundary conditions at the centre and surface of the star \citep[e.g.,][]{Kippenhahn}. The former are defined by the requirement
 that $m = 0$, $L = 0$ at $r = 0$, while for the latter, an atmospheric model is required. For the latter, we will use a grey atmospheric model, based on a $T-\tau$ relation described by the Hopf function modelled on \cite{Vernazza1981}. For the hydrostatic equilibrium in the atmosphere we assume standard gravity, since the effect of the fifth force in that region is irrelevant.  

The fifth-force term in \eqn~(\ref{eq:dP_dr}) depends on the second derivative of the mass profile with respect to radius, which we can express as
\be
\frac{\dd ^{2} m(r)}{\dd r^2} = \frac{\dd m(r)}{\dd r} \left( \frac{\dd  \log \rho}{\dd r} + \frac{2}{r} \right), \label{eq:d2mass}
\ee
which is the form which we also adopt for the numerical implementation. Note that in the above equation $\dd m/\dd r = 4 \pi \rho(r) r^2$. The derivatives in \eqn~(\ref{eq:d2mass}) are implemented in a discreet manner as \be
\frac{\dd  \rho(m)}{\dd m} \simeq \frac{\rho_{k+1} - \rho_k}{ m_{k+1} - m_k }, 
\ee
where $k$ labels the $k$-th cell on the grid and all physical variables are evaluated either at the centre or the edge of the $k-$th cell. The fifth force is then evaluated at each time step ($t_{k}$) next to the standard Newtonian term, from the outermost ($k = 1$) to the innermost cell corresponding to the stellar surface and core respectively. For a typical run, we use about $~\sim 1000$ spatial grid points and a time step of $\sim10^7$ years. 

Our microphysics assumptions are as follows. For the EoS we use OPAL \citep{Rogers2002}, whereas the opacities are based on the tabulations from OP \citep{Seaton2005}
with low- and high-temperature extensions from \citet{Ferguson2005} and \citet{Buchler1976} respectively. Nuclear and weak reaction rates are from JINA REACLIB \citep{Cyburt2010}  and \citet{Fuller1985, Oda1994, Langanke2000} respectively. We also choose to set overshooting to zero, an assumption which does not restrict the generality of results. In particular, some of our key results presented here we confirmed allowing for a small (exponential) overshooting. An important input is the initial fraction of metals, which affects the computation of opacities too. Unless otherwise stated, we choose to work with the so--called GS98 fraction based on \citet{GS98}, due to their good agreement with helioseismic data. Later on, we discuss the significance of this choice in the light of the revised solar metallicity of \cite{A09} (A09) and the abundance problem. 

\begin{table}
\begin{tabular}{l*{6}{c}r} \label{table:models}
Metals mixture   & $\YMG$ &   $Y^{\text{initial}}$ &  $Z^{\text{initial}}$  & $\alpha_{\text{MLT}}$    \\
\hline
\hline
GS98        & $0$               & 0.2690       &           0.0186            &   2.0031     \\
GS98 (Free EoS)& $0$     &0.2690       &             0.0186            &   2.0017     \\
A09           & $0$               & 0.2699     &           0.0159             &  2.0732       \\
GS98        & $0.01$      & 0.2670      &              0.0187             &   1.9962      \\
GS98        & $-0.01$     &0.2709       &             0.0185            &   2.0075     \\
\end{tabular}
	\caption{\label{table:models}
	A description of solar models used in our analysis, and their initial parameters. All are calibrated to reproduce the observed solar surface metallicity, radius and effective temperature within $1 \sigma$ of the respective error (see also Table \ref{table:observables}), and have standard (unmodified) opacity and diffusion coefficients. They are computed with our reference EoS (OPAL EoS), except the model computed with the alternative choice of Free EoS. The first model is our reference model at standard gravity and standard input physics (opacity, diffusion, etc.), and is based on the ``old" metallicity mixture of \cite{GS98}, which is in good agreement with helioseisimic observations. Model A09 uses the revised metallicity of \cite{A09}, which is, however, in tension with helioseismic data (see Fig.~\ref{fig:calibration}). The latter model is not used for computations with the fifth force, but was computed for comparison with GS98. The models in the table computed with the fifth force turned on ($\YMG =\pm 0.01$) are used for illustration purposes throughout the text. More details about our choice of input physics are provided in Section~\ref{sec:calibration}. }
\end{table}

\begin{table}
\begin{tabular}{l l }
Observable            & { Value $\pm 1\sigma$} \\   
\hline
\hline
\vspace{0.1cm}
Solar radius [$R_{\odot}$]                 &   $1 \pm 10^{-4} $     \\
Effective temperature, $T_{\text{eff}} [K]$    & $5777  \pm 66 $       \\
Surface abund., $(Z/X)_{\text{surf.}}$ (GS98)     &  $0.02292 \pm 10^{-3}$     \\
Surface abund., $(Z/X)_{\text{surf.}}$ (A09)        & $0.0178 \pm 10^{-3}$    \\
	Base of conv. zone, $R_{cz} [R_{\odot}]$     &  $0.713 \pm 10^{-3}$    \\
\end{tabular}
	\caption{\label{table:observables}A summary of the solar observables and their precision, used in this work. The quoted value for the radius of the base of the convective zone, $\Rcz$, is the one inferred from helioseismic observations \citep[e.g.,][]{JCD:1991, basu1997}. The old (GS98) and the revised (A09) surface metallicity is according to \citet{GS98} and \citet{A09} respectively.}
\end{table}

\begin{figure}
\includegraphics[width=80mm]{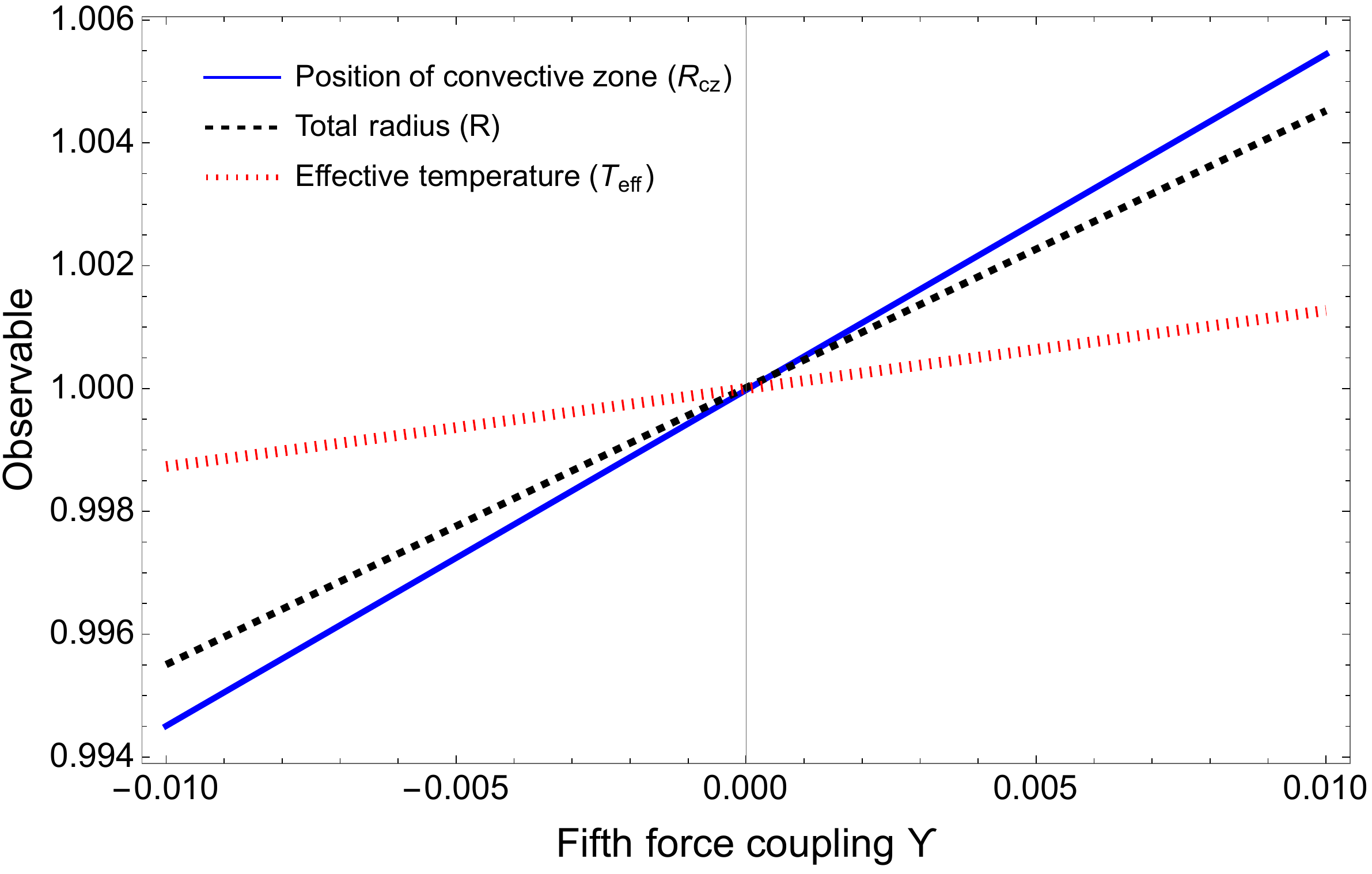}
	\caption{\label{fig:uncalvar} Variation of total radius, base of convective zone and effective radius under the fifth-force for {\it uncalibrated} solar models. Each observable is normalised with respect to its value at $\YMG = 0$ (no fifth force) at a fixed initial helium abundance ($\Yin$), initial metallicity ($\Zin$) and mixing-length parameter ($\aMLT$). Whereas the fifth force has a significant effect on the radius and position of the convective zone, its effect is milder on the effective temperature. One notices that the weakening of gravity for most part of the star, i.e $\YMG > 0$, leads to an increase in the total radius and vice versa. Similarly, a weakening of gravity further tends to make the convective zone shallower. }
\end{figure}
\begin{figure}
 \includegraphics[width=80mm]{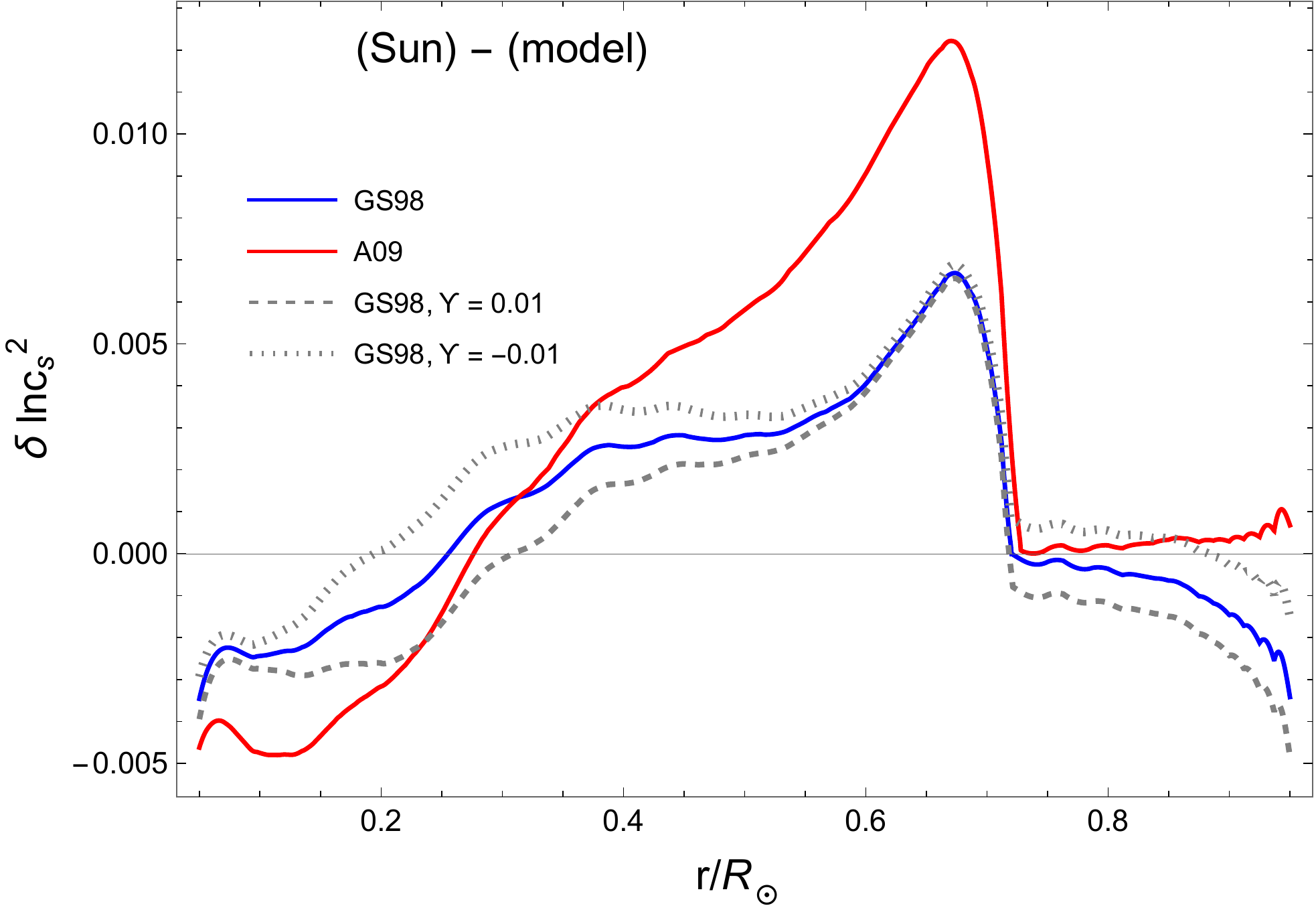}  \\
	\caption{ \label{fig:calibration} Fractional difference for the sound speed in the sense (Sun) - (Model). ``Sun" refers to the sound-speed profile as inferred from helioseismic observations at standard gravity according to the results of \cite{inversion_data}. Models GS98 and A09 correspond to theoretically computed {\it calibrated} solar models at standard gravity ($\YMG = 0$) with the two distinct, and most popular metallicity mixtures for the Sun (see also Table \ref{table:models}, and Section~\ref{sec:calibration}). The discrepancy between the predictions of the GS98 and A09 model, when compared to helioseismic observations, is what is known as the solar abundance problem -- if the metallicity measurements of A09 are assumed to be correct, the discrepancy is typically attributed to a mismatch in the modelling of opacities in the solar interior. For comparison, we also show two computed models with the fifth force turned on  ($\YMG = \pm 0.01$) based on the GS98 mixture. We note that the fifth-force effect peaks in the radiative zone (see also Figure~\ref{fig:model_differences}) which will be key for constraining the fifth force with helioseismic observations. }
\end{figure}

\subsection{Solar evolution simulations}
It is important to first lay out our framework for testing gravity with the Sun. Solar observables such as luminosity and surface temperature are known to very high precision, whereas thanks to helioseismic observations the same holds for the interior solar profiles such as density and speed of sound. Our approach to test the fifth force is as follows: We choose to work with {\it calibrated} solar models, i.e models which match the equilibrium observables of the Sun within their $1\sigma$ errors. Any effect of the fifth force will then impact the structure of the interior solar profiles, and the latter will be our key quantities for investigation. 

The computation of calibrated solar models corresponds to finding that subspace of initial conditions which reproduce the observables of the present Sun. Our main present-day observables are the total stellar radius ($R$), surface element abundances ($(Z/X)_{\text{surf.}}$) and luminosity ($L$) (see also Table \ref{table:observables}). The latter can be traded for the effective temperature $T_{\rm{eff}}$ given the radius $R$.
In addition, we may include results of helioseismic analyses, such as the radius $R_{\rm cz}$ at the base of the convection zone, the envelope helium abundance, or the inferred solar sound speed. Our initial parameters correspond to the mixing-length parameter ($\alpha_{\text{MLT}}$), initial helium ($\Yin$) abundance and initial metallicity ($\Zin$). On top of the previous parameters, we need to define the value of the fifth-force coupling $\mathcal{Y}$. Therefore, our parameter space is 4-dimensional. We choose to fix our initial conditions at the stage of the zero-age main sequence, from which point the star is left to evolve until today. A calibrated solar model corresponds to a model which {\it simultaneously} lies within the statistical tolerance of the observables. 

We start with a solar calibration without fifth force, i.e. $\mathcal{Y} = 0$. This enables us to pin down the 3-dimensional parameter space of solar parameters before we turn on the fifth force, and will produce our base model at standard gravity. We employ the simplex method, which successively seeks for the values of the initial parameters ($p_i$) minimising a $\chi^2$ functional built out of the desired observables ($O_i$),
\be
\chi^2 =   \sum_i  \frac{ \left[ O_i(p_j) - O_i(\text{Sun}) \right]^2}{\sigma_{i}^2}. \label{eq:chi2}
\ee
Typically, for a solar calibration, 
$$
p_i = \{ \aMLT, \Yin, \Zin\}, \; \; \;
 O_i = \{ \Teff, \Rs, \ZX \} \; .
$$ 
Though this calibration approach is conceptually different from the traditional method of linearisation, the two methods should coincide modulo small differences of numerical origin. We have verified this agreement between the two procedures for indicative solar calibrations. In our computations we employ the GS98 metallicity mixture, but for the sake of comparison we also calibrate a model based on the revised abundance A09 \citep{A09}. (For a discussion on the solar abundance issue see section \ref{sec:metallicity_problem}.) The set of representative calibrated solar models we use in our analysis are summarised in Table \ref{table:models}. 

We proceed turning on the fifth-force term using our GS98 calibration as our reference model, which extends our parameter space to
$$p_i = \{ \aMLT, \Yin, \Zin, \YMG\}.$$
Given that we work with calibrated models, the effect of the fifth force will manifest itself in the interior profiles. Some useful intuition on the effect of the fifth force on a calibrated model can be drawn from homology arguments. In this context, the luminosity is related to the total mass, radius and mean molecular weight through \citep{Kippenhahn},
\be
L \propto \frac{G^{15/2} \mu^{15/2} M^{11/2} }{R^{1/2}}. 
\ee
At fixed mass and radius, and for $\YMG > 0$, if the luminosity $L$ is to remain the same, the strengthening of gravity in the solar core (see Figure~\ref{fig:delta_Geff}) requires the compensating effect of a decreased $\mu$. Figure~\ref{fig:uncalvar} shows the effect of fifth force on some solar observables, for uncalibrated models. One see that the fifth force predominantly affects the total radius and the thickness of the convective envelope, and to a smaller extent temperature.

\begin{figure*}
\begin{tabular}{cc}
\hspace{3.5cm} \includegraphics[width=95mm]{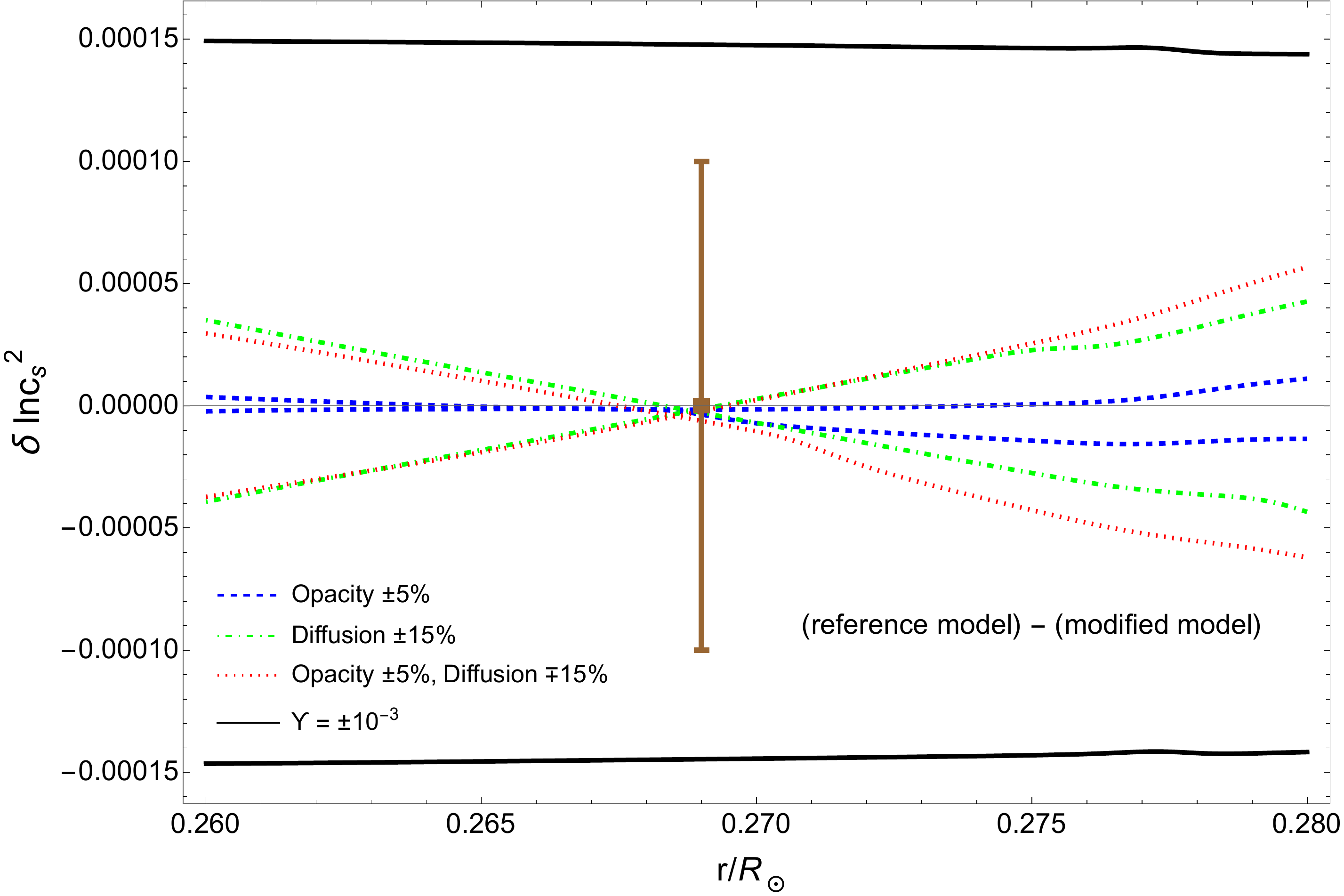}   \\
\hspace{-4cm}  \includegraphics[width=90mm]{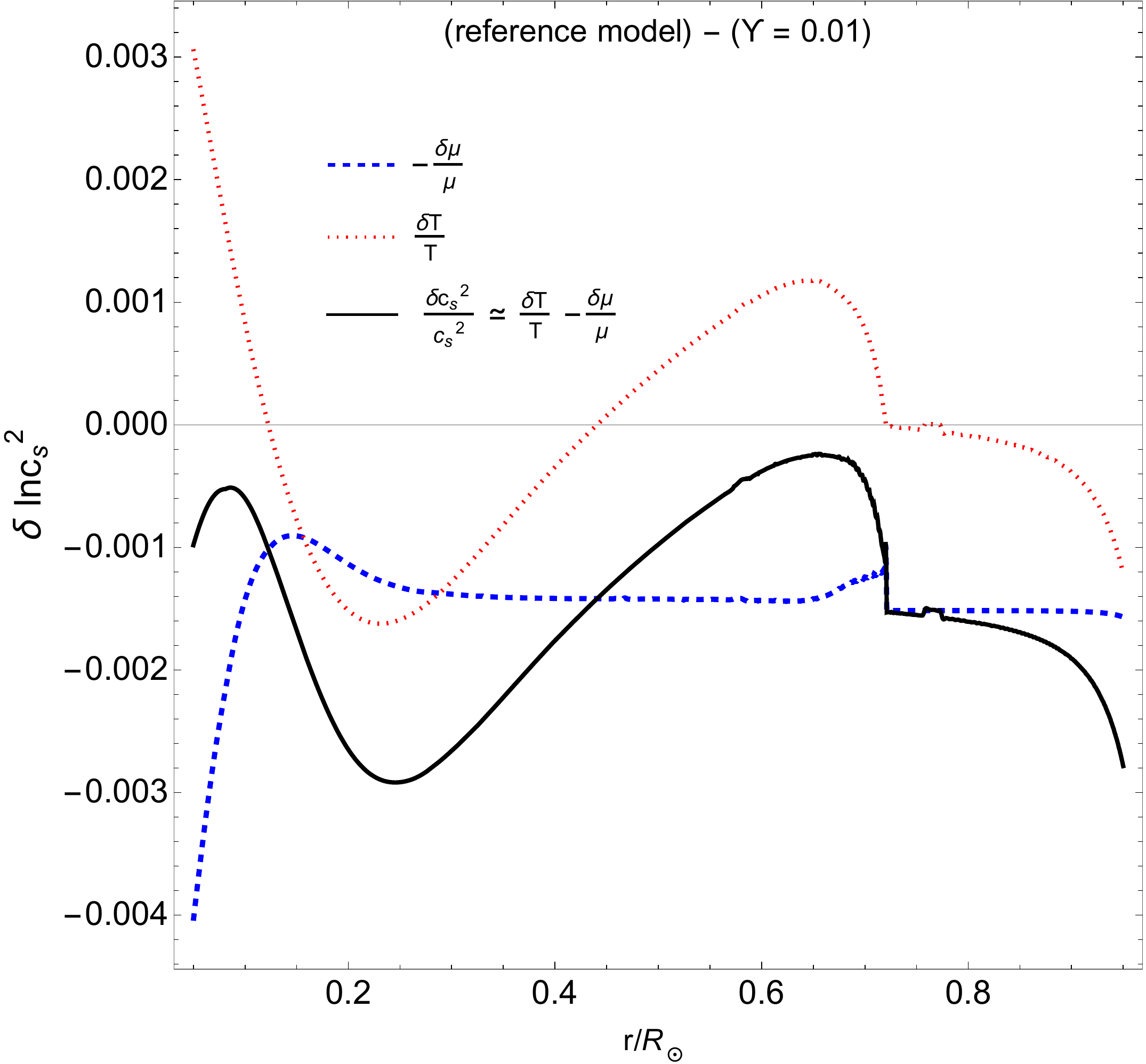} &  \hspace{-4cm} \includegraphics[width=90mm]{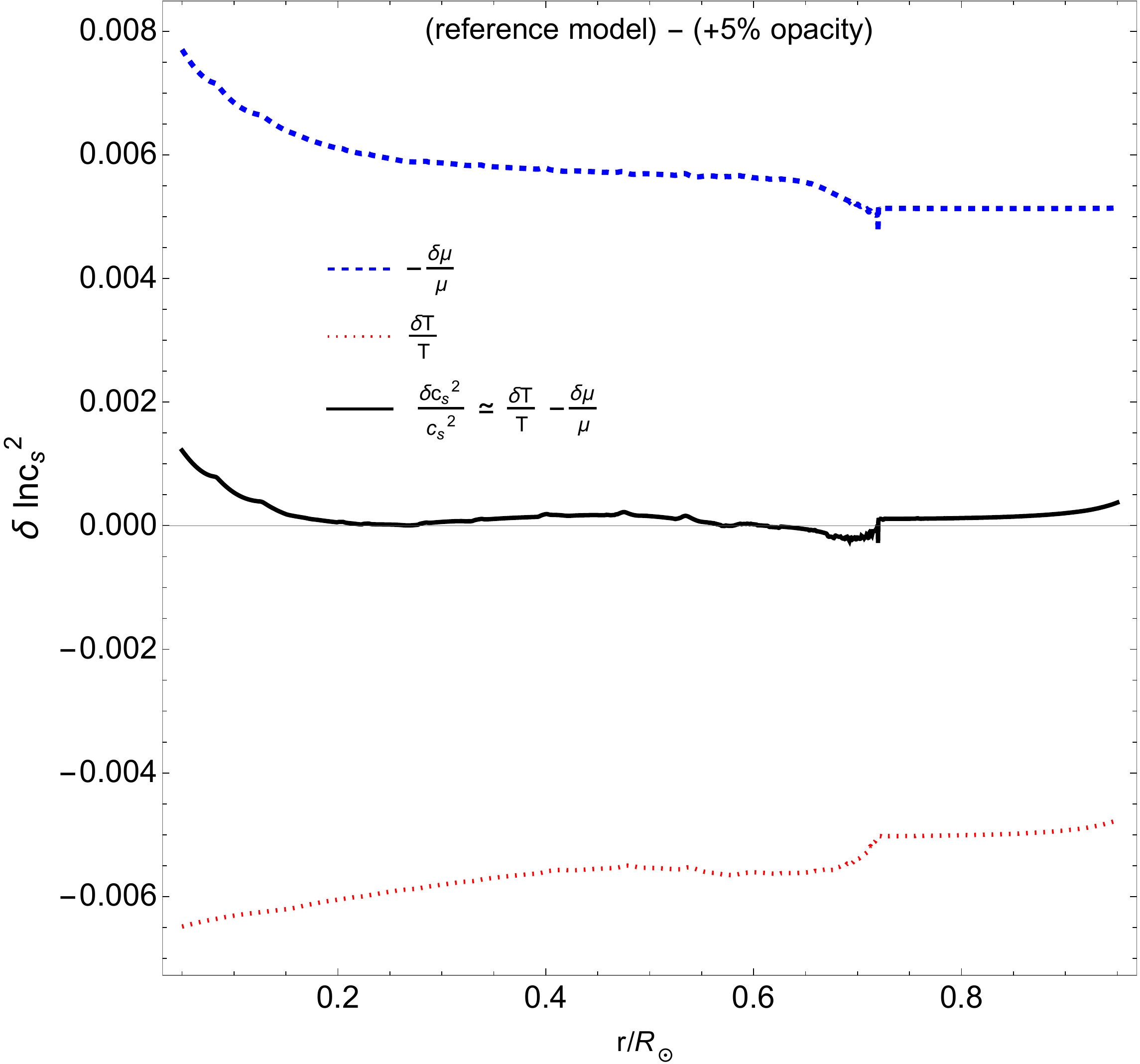} 
\end{tabular}
\caption{\label{fig:delta_cs_components} 
	{\bf Top}: Model differences for the sound-speed profile similar to those of Figure~\ref{fig:model_differences}, in the sense (reference model) - (modified model), but with a smaller fifth-force coupling $\YMG = 10^{-3}$. The plot zooms into the solar region where our estimates of systematics from opacity and diffusion may become negligible compared to that of the fifth force, in order to clearly illustrate the peak of the fifth-force effect in the radiative zone. The error bar shows approximately the error of helioseismic inversions at that point, $\sim 10^{-4}$, to highlight the precision down to which helioseismic reconstructions of the solar sound speed can constrain the coupling $\YMG$. The insignificance of the opacity/diffusion uncertainties compared to the fifth force for this solar region is discussed in Section~\ref{sec:systematics} and further illustrated in Figure~\ref{fig:model_differences}. \\
{\bf Bottom}:
	An illustration of the mechanism behind the ``clean" peak of the fifth-force effect on the speed of sound profile around $r \simeq 0.25 R_{\odot}$, where the systematics from uncertainties related to opacity and diffusion become negligible compared to the effect of the fifth force (see also Figure~\ref{fig:model_differences}). The reference model at standard gravity is the GS98 model (see Table \ref{table:models} and text). The plots assume for illustration the ideal-gas approximation for the sound-speed profile, where $c_{s}^2 \propto \Gamma_{1} T/\mu$, and neglect the small variation of the adiabatic index $\Gamma_1$ (see also Figure~\ref{fig:model_differences}). The right panel shows the fractional difference of the sound-speed profile and its components under an increase of the opacity by $5\%$, but the picture is similar for uncertainties in diffusion. The almost zero variation of the sound-speed profile around  $r \simeq 0.25 R_{\odot}$, i.e $\delta c_{s}^2 \simeq 0$, is due to the fine cancellation between the contribution of the mean molecular weight and temperature. The left panel shows how the picture changes for the fifth-force effect -- the induced change in the mean molecular weight under the fifth force, thanks to the change in the hydrogen fraction, in combination with the non-trivial scaling of the temperature profile, leads to the domination of the fifth-force effect compared to opacity/diffusion uncertainties in this solar region. }
\end{figure*}

\begin{figure*}[h!]
\hspace{-1.0cm} \begin{tabular}{cc}
  \includegraphics[width=90mm]{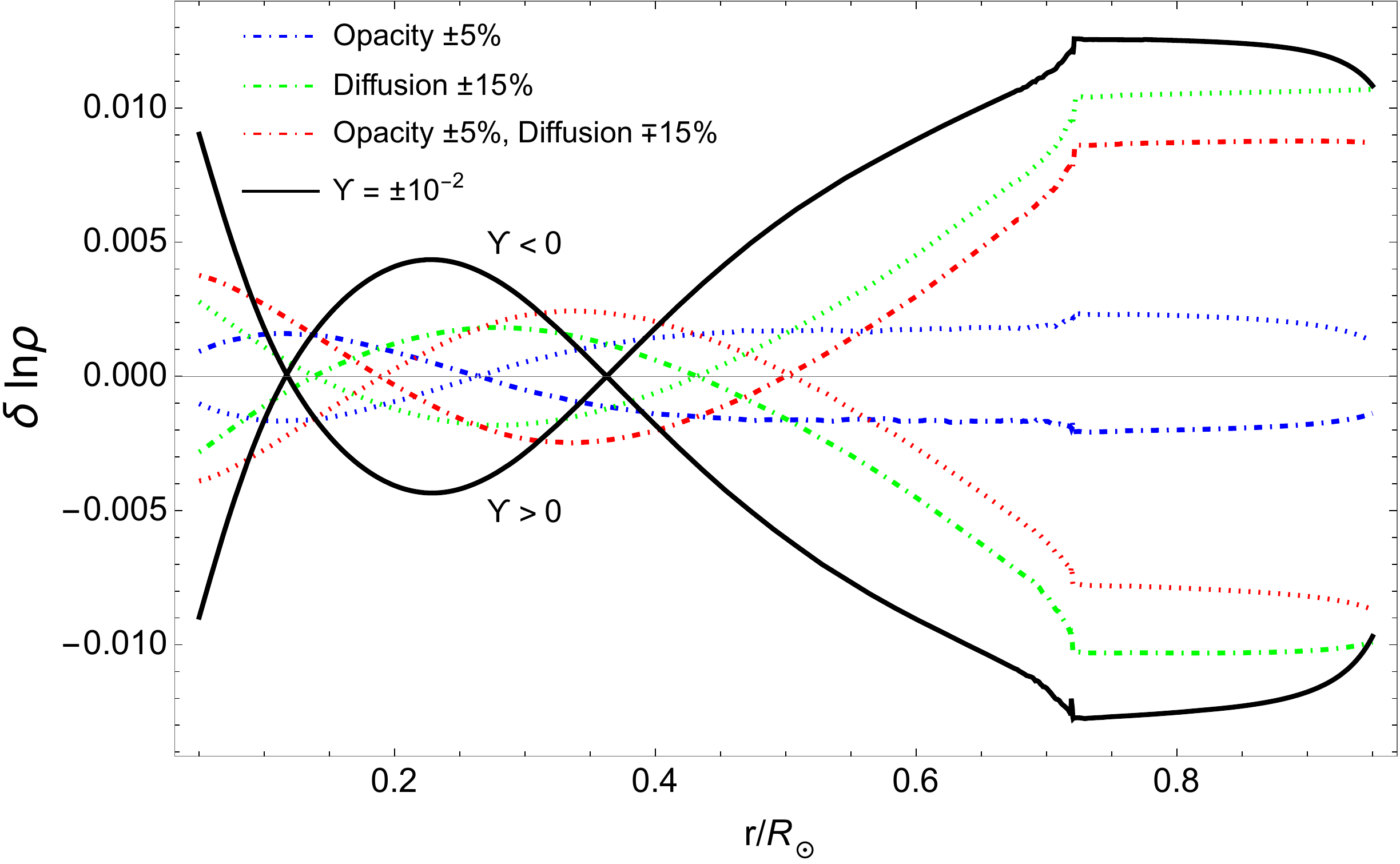} &   \includegraphics[width=90mm]{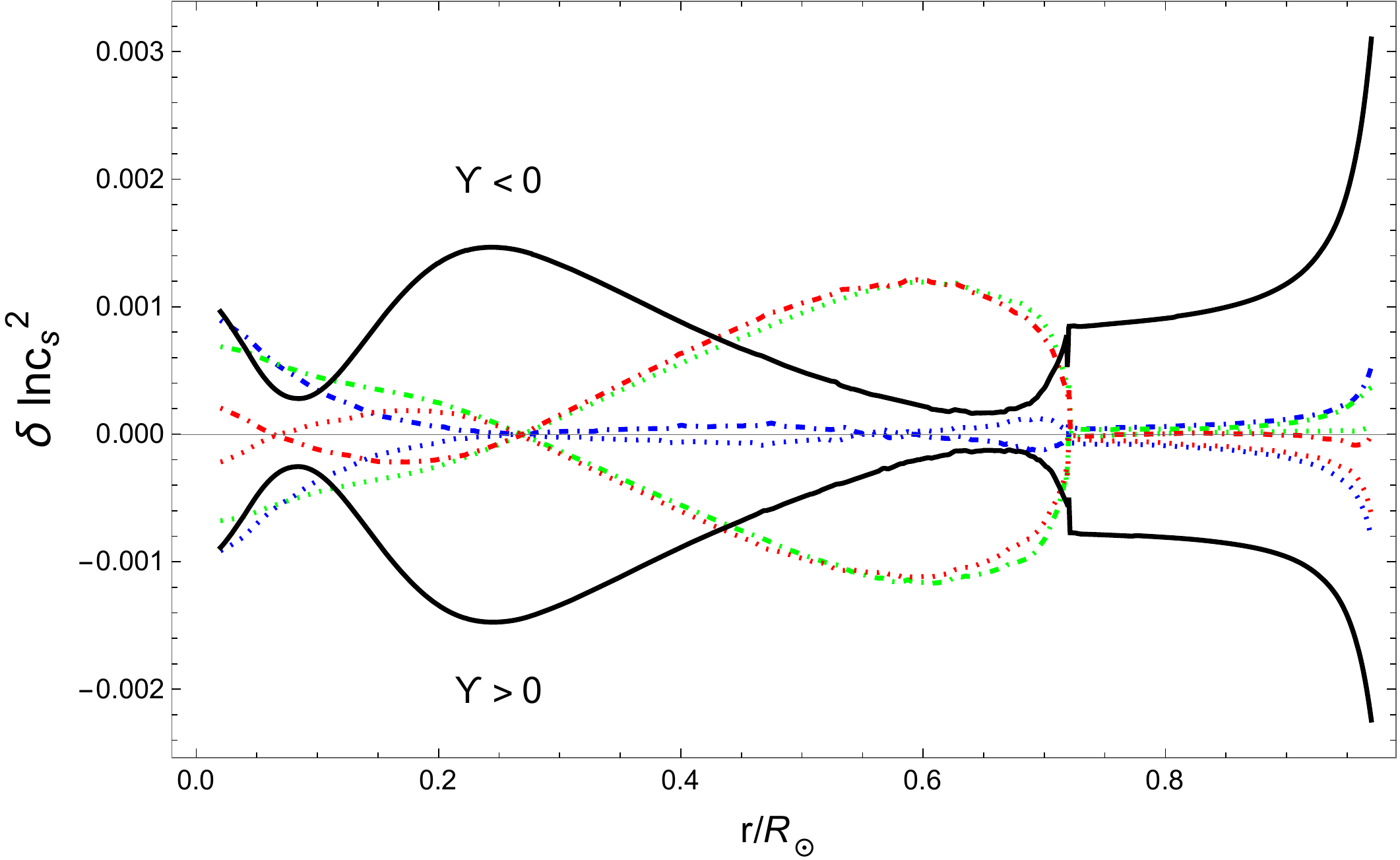} \\
(a) Density  & (b) Speed of sound squared  \\[6pt]
 \includegraphics[width=90mm]{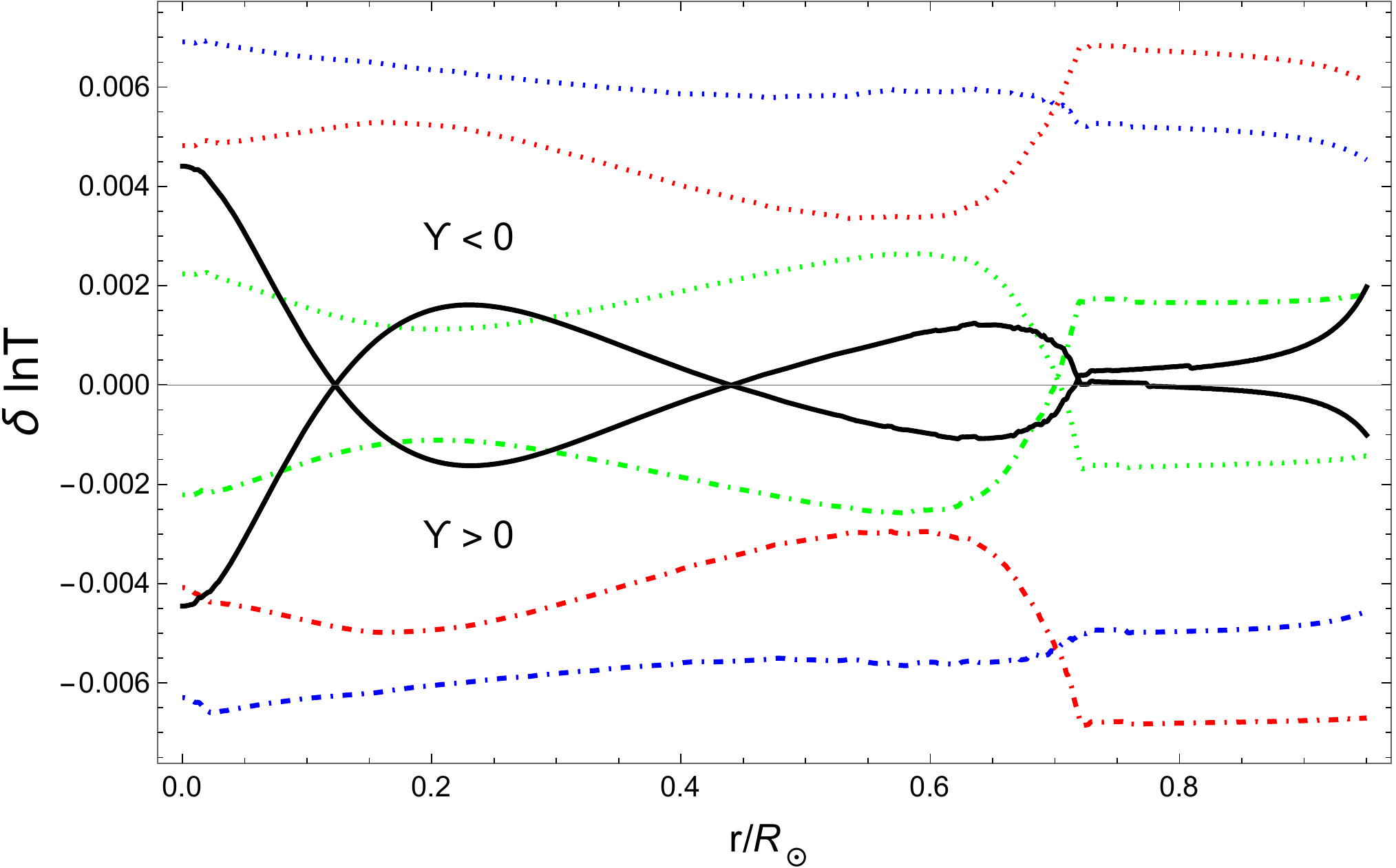} &   \includegraphics[width=90mm]{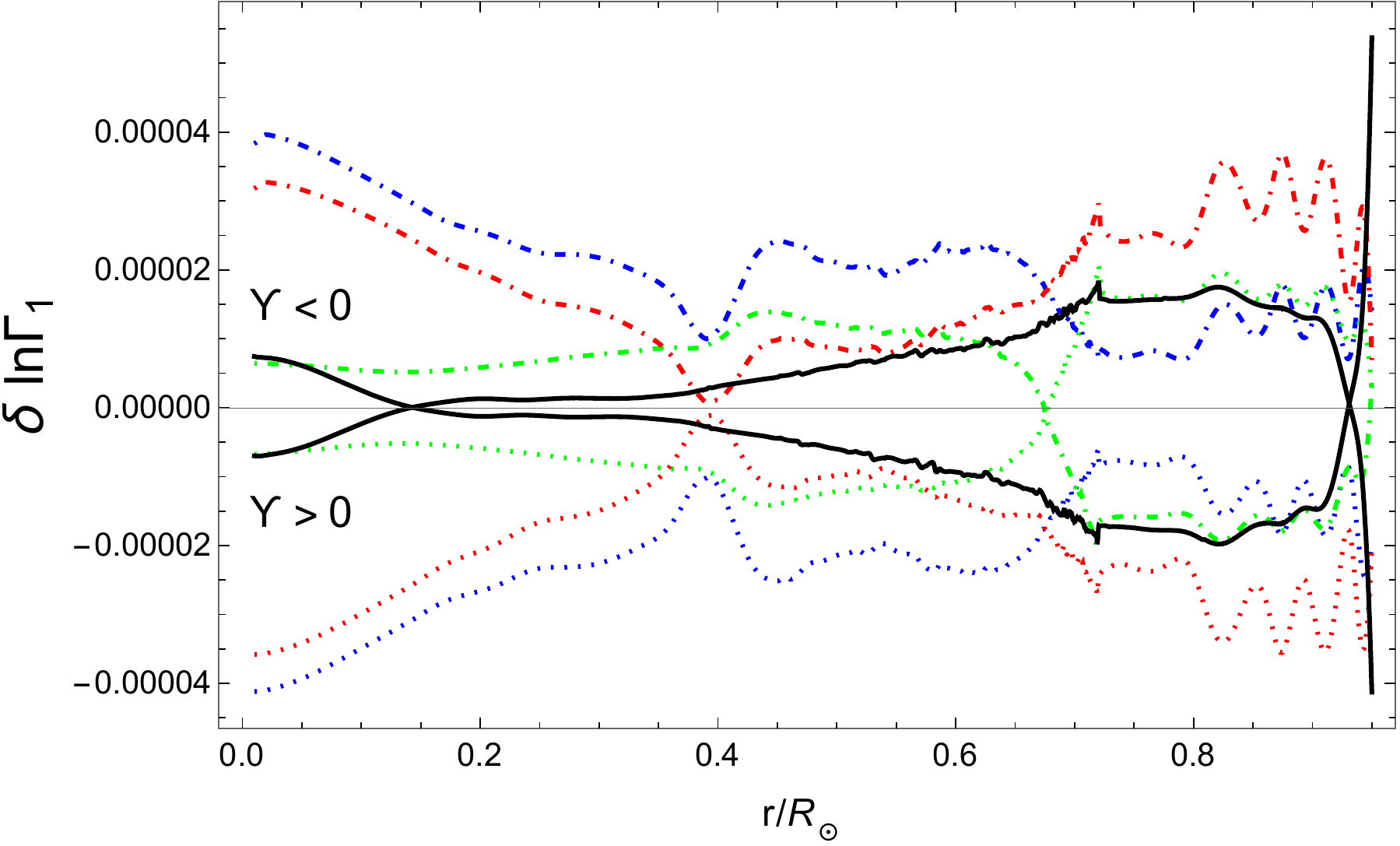} \\
(c) Temperature  & (d) Adiabatic index  \\[6pt]
 \includegraphics[width=90mm]{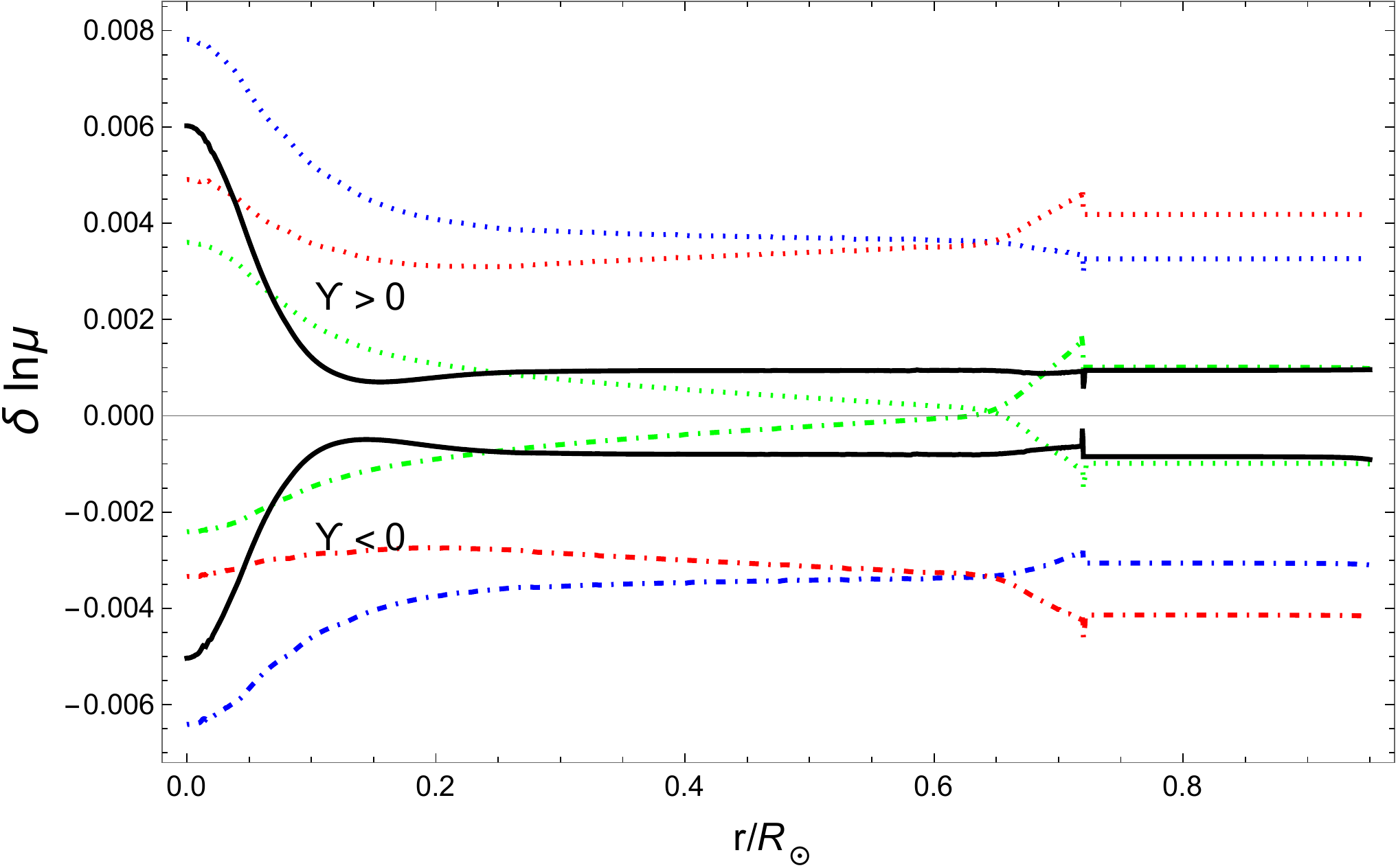} &   \includegraphics[width=90mm]{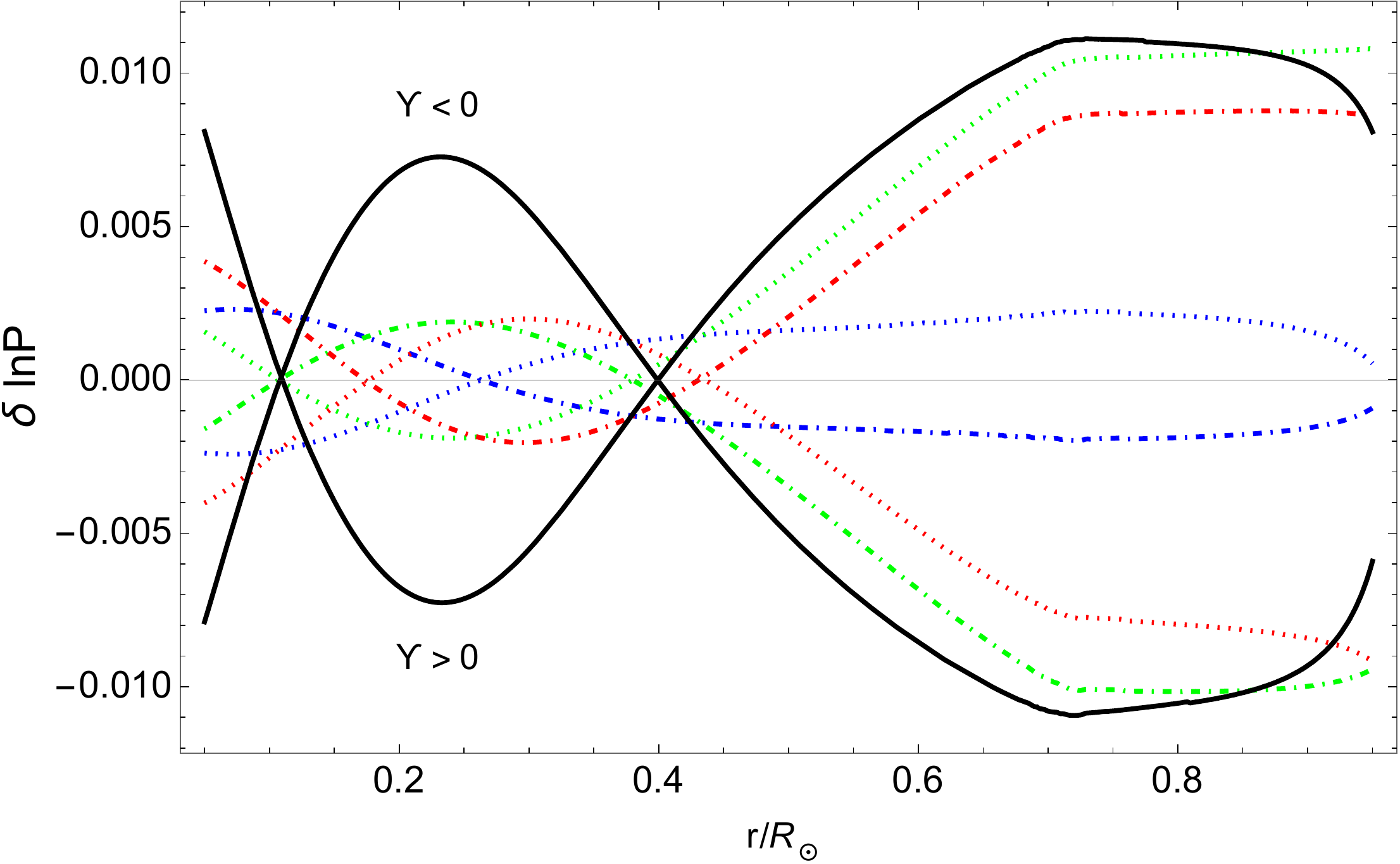} \\
(c) Mean molecular weight  & (d) Pressure  \\[6pt]
\end{tabular}
	\caption{ \label{fig:model_differences} Fractional model differences in the sense (reference model) - (modified model) for solar interior profiles, to illustrate the importance of the most significant uncertainties in solar modelling (opacity, diffusion) against the fifth-force effect. The legend for the curves is shown in the top left panel. The dot-dashed (dotted) curves correspond to an increase (decrease) of the respective input physics parameter. For the combined change of opacity/diffusion the dot-dashed curves correspond to increased opacity and decreased diffusion.  The reference model is the calibrated solar model at standard gravity with GS98 metallicity and with unmodified opacity and diffusion coefficients, as explained in the text and Table \ref{table:models}. In all panels, the models with the fifth force turned on (black, solid curves) are computed on top of our reference model at standard gravity, and have a value of the fifth-force coupling $|\YMG| = 10^{-2}$ for illustration purposes.  As conservative estimates for the uncertainties in input physics we use $\pm 5\%$ for opacity and $\pm 15\%$ for diffusion, which are computed on top of our reference model (see Section~\ref{sec:systematics}). We also consider the combined effect of the simultaneous uncertainty of opacity and diffusion. The mild variation of the density and sound-speed profiles around the base of the convective zone ($\simeq 0.71 R_{\odot}$) has been explained through analytic relations in Section~\ref{sec:theory}.}
\end{figure*}

\section{Input physics systematics and fifth force} \label{sec:systematics}
Searching for a departure from Newtonian gravity in the Sun is a rather delicate task, due to interplay between the well-observed solar structure and the small fifth-force effect. Therefore, an understanding of the systematics that could hinder a clean extraction of the fifth-force effect is necessary. In this section, we will use our numerical results to attack this issue. We will show that the effect of the fifth force leaves a characteristic and clean signature on the solar sound-speed profile which allows it to be accurately distinguished from typical theoretical uncertainties in the modelling of the Sun. 

\subsection{Opacity and diffusion}

Realistic opacity profiles for the Sun have a highly non-trivial dependence on the density, temperature and element abundances, and they contribute to one of the most important theoretical uncertainties in solar modelling. For example, an increase in opacity at standard gravity, will tend to increase the density in most parts of the star, an effect that can be also achieved by strengthening gravity at constant opacity. The uncertainty of opacity modelling in the Sun ranges from $\sim 2\%$ near the centre of the Sun, to an increase of up to $7\% \pm 4\%$ around the base of the solar convective zone, where element ionisation becomes significant. The $2\%$ uncertainty close to the centre is typically computed as the averaged opacity difference between the two popular opacity tabulations, namely the OPAL \citep{Iglesias1993,Iglesias1996} and OP opacity tables. The $7\%$ uncertainty results from a comparison of opacity tables with a measurement of iron's opacity at the laboratory. Here, as a conservative estimate, we adopt a constant error of in the opacity of
\begin{equation}
\kappa = \kappa_{0} \pm 5\%,
\end{equation} 
with $\kappa_0$ the reference opacity profile based on the OP tabulation, and the chemical mixture GS98 (see also Table \ref{table:models}). We implement this uncertainty by increasing/decreasing the opacity throughout the star by the same constant factor, $1 \pm 0.05$. 
 
The diffusion error is defined as the theoretical uncertainty in the diffusion coefficients for the various elements. The latter govern the efficiency of diffusion in the star, and in the context of the diffusion treatment of \cite{thoul94} employed in our simulations, the comparison between different solutions for the diffusion equations can be used to place a conservative uncertainty of $15 \%$. We will therefore use for the uncertainty on diffusion,
\begin{equation}
D_i = D_i \pm 15 \%,
\end{equation}
where $D_i$ the diffusion coefficients related to element $i$. Notice that the uncertainty is applied the same to the coefficients of all elements. For a further discussion on the topic we refer to \citet{Vinyoles:2016djt} and references therein. 

The effects of variations in opacity and diffusion (at standard gravity), as well as that of the fifth force (at standard opacity/diffusion), are shown in the model differences of Figure~\ref{fig:model_differences}.
Here the effect on the thermodynamic properties of the change in composition is illustrated in terms of the mean molecular weight, approximately calculated as
\be
\mu = {4 \over 3 + 5X - Z}.
\ee

In addition to the individual effects of opacity and diffusion, we also show the combined effect resulting from the simultaneous variation of both. At standard gravity, an increase in opacity will generally tend to increase the density, temperature and mean molecular weight, and decrease the sound speed. A similar situation holds for an increase in the efficiency of diffusion, but in that case, the sound-speed profile decreases. Now, for a fifth-force coupling $\YMG > 0$, density and pressure will generally tend to decrease for most of the solar region, while the sound speed will decrease. The opposite situation holds for $\YMG < 0$. The increase of density, pressure and temperature before the turn over at $r \simeq 0.2 R_{\odot}$ in Figure~\ref{fig:model_differences}, is due to the increase of gravity in that region, as follows from the scaling in \eqn~(\ref{fig:delta_Geff}). 

One also notices that the change in sound speed in the convection zone resulting from changes in opacity or diffusion is small 
compared to the changes in density or pressure. Within the polytropic solutions, \eqns~(\ref{eq:rho0-solution}) and (\ref{eq:c^2-convective}),
the structure generally depends on the constant $K$ which is different for each modified model.
This directly affects density (see \eqn~(\ref{eq:rho0-solution})) and hence pressure, while \eqn~(\ref{eq:c^2-convective}) shows that the sound speed is independent of $K$.
In contrast, \eqn~(\ref{eq:delcsq}) shows that the fifth force directly affects the sound speed in the convection zone, leading to the behaviour seen in Figure~\ref{fig:model_differences}.

The peak around $r = 0.25 R_{\odot}$ in the fifth-force effect on density, pressure and sound speed shown in Figure~\ref{fig:model_differences} was first observed in \cite{Saltas:2019ius} under the crude approximation of an $n=3$ polytrope. However, the striking feature shown here is that it not only persists within our accurate simulations, but it occurs at a point where the systematics from opacity and diffusion become negligible compared to it. To understand this better, we can use the ideal gas approximation for the EoS (see \eqn~(\ref{eq:cs-frac-ideal1})). Then,
\begin{equation}
\frac{\delta  c^{2}_s}{ c^{2}_s} \simeq  \frac{\delta T}{T} - \frac{\delta \mu}{\mu}, \label{eq:cs-ideal-frac}
\end{equation}
where we neglected the insignificant variation of $\Gamma_1$ under model differences compared to the other terms (see also the model differences of Figure~\ref{fig:model_differences}). Under variations of opacity and/or diffusion at standard gravity, the contributions coming from the variation of temperature and mean molecular weight approximately cancel with each other around the point $r \simeq 0.25 R_{\odot}$. This is not anymore true for model variations under the fifth force (at unmodified opacity/diffusion), where the temperature traces the change of the effective strength of gravity with radius in that region (see also \eqn~(\ref{fig:delta_Geff})). An illustration of this mechanism is shown in Figure~\ref{fig:delta_cs_components}. This surprisingly clean signal of the fifth force allows for precision tests of the theory with helioseismic observations. The comparison of indicative computed models under the fifth force against helioseismic data is shown in Figure~\ref{fig:delta-cs-real-syst}.  

\begin{figure}
 \hspace{-0.4cm}\includegraphics[width=90mm]{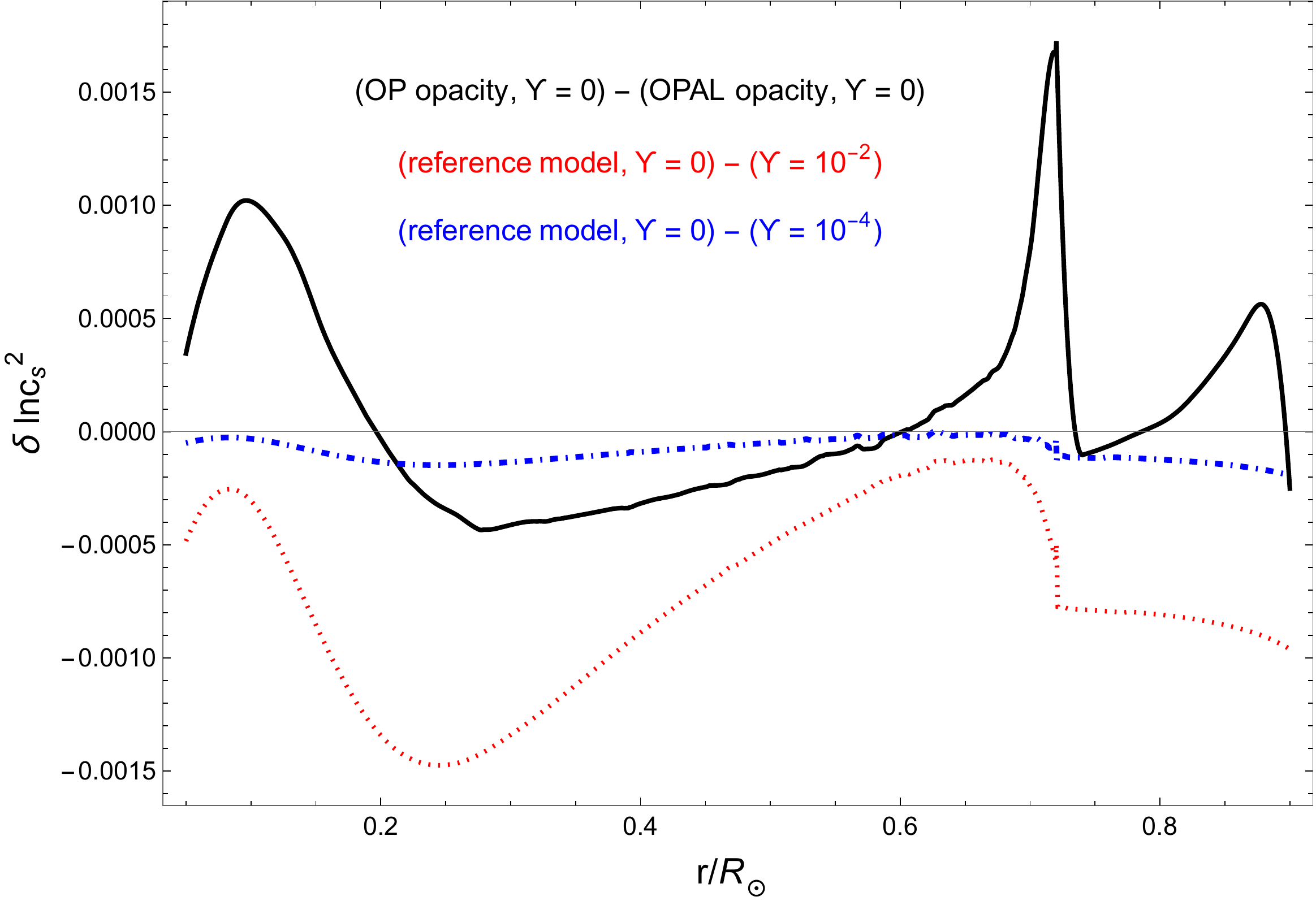}
\caption{ \label{fig:deltacs-opacity} The difference between solar profiles computed with different opacity tabulations provides an estimate of the error coming from opacity modelling. Here, we show the difference in the sound speed for a model computed with our reference opacity table (OP) and a model computed with an alternative opacity (OPAL). Both computations assume standard gravity ($\YMG = 0$), and are based on the GS98 metallicity mixture. In our analysis of systematics, the uncertainty in the modelling of opacity was considered constant throughout the star, as explained in Section~\ref{sec:systematics}. However, one cannot exclude the possibility of a localised opacity variation such as the one emerging from the comparison between OP and OPAL opacity tabulations. The localised opacity variation around $r \simeq 0.3 R_{\odot}$ could interfere with the signature of the fifth force on the sound-speed profile. For comparison purposes we also show the corresponding impact of the fifth force on the sound speed computed with our reference opacity (OP) and for two values of the fifth force coupling. We discuss the impact of this on inferences of the fifth-force coupling $\YMG$ in Section \ref{sec:constraint}. }
\end{figure}

Before we close this section, we want to challenge our assumption of a constant uncertainty in the opacity throughout the star. In particular, we ask what would be the effect of a radius-dependent uncertainty which would peak around the region where the fifth-force effect peaks. We approach this by looking at the difference between two popular opacity tabulations, namely OPAL and OP, which provides an estimate of the opacity error. The model differences between the latter opacity tabulations have been previously studied in the literature (at standard gravity), see e.g. Figure~$30$ of \citet{JCD:2021}, and it is illustrated in Figure \ref{fig:deltacs-opacity}. The induced difference in the sound-speed profiles computed with OPAL and OP, respectively, shows a peak at $r \simeq 0.3 R_{\odot}$, inducing a reduction of the sound speed which can reach up to $0.2 \%$. Such a localised opacity uncertainty  could interfere with the fifth-force effect and we comment on it further in Section \ref{sec:constraint} in light of new constraints on $\YMG$. A detailed confrontation of different opacity tabulations against helioseismic data has been performed in \cite{Villante2014} where it was found that standard solar models computed with the OP opacity do not provide a good fit to the sound-speed profile as inferred from helioseismology in the region $0.3 < r/R_{\odot} < 0.6$. We stress though, that such analyses would have to be in principle revisited in the presence of the fifth force, since one now has one extra degree of freedom to be accounted for. 
Furthermore, opacity-related uncertainties are also related to uncertainties in the metallicity fraction. As we discuss in Section \ref{sec:metallicity_problem} (see also Figure \ref{fig:calibration}) the proposed metallicity mixtures A09 shows tension with helioseismology, although the very recent analysis of \cite{magg2022} suggests that the tension is significantly settled through a revised mixture, close to the GS98 one.

\subsection{Equation of state}
The EoS affects the interior profiles, including the sound speed,  in two different ways. One is the direct effect of different input physics on the adiabatic index $\Gamma_1$, as the latter provides a proxy on the way the pressure relates to density with radius. The effect of the fifth force on the adiabatic index is shown in Figure~\ref{fig:model_differences}. It is evident that fifth-force effect on $\Gamma_1$ is negligible throughout the Sun, reaching a maximum of about $0.002\%$ in the convective zone, where it is comparable to the respective effect of diffusion. It should be also noted that for most of the radiative zone its effect is smaller than the effect of opacity and diffusion.
However, the EoS also determines the relation between pressure, density and temperature and hence the overall structure of the model.

\begin{figure}
 \includegraphics[width=90mm]{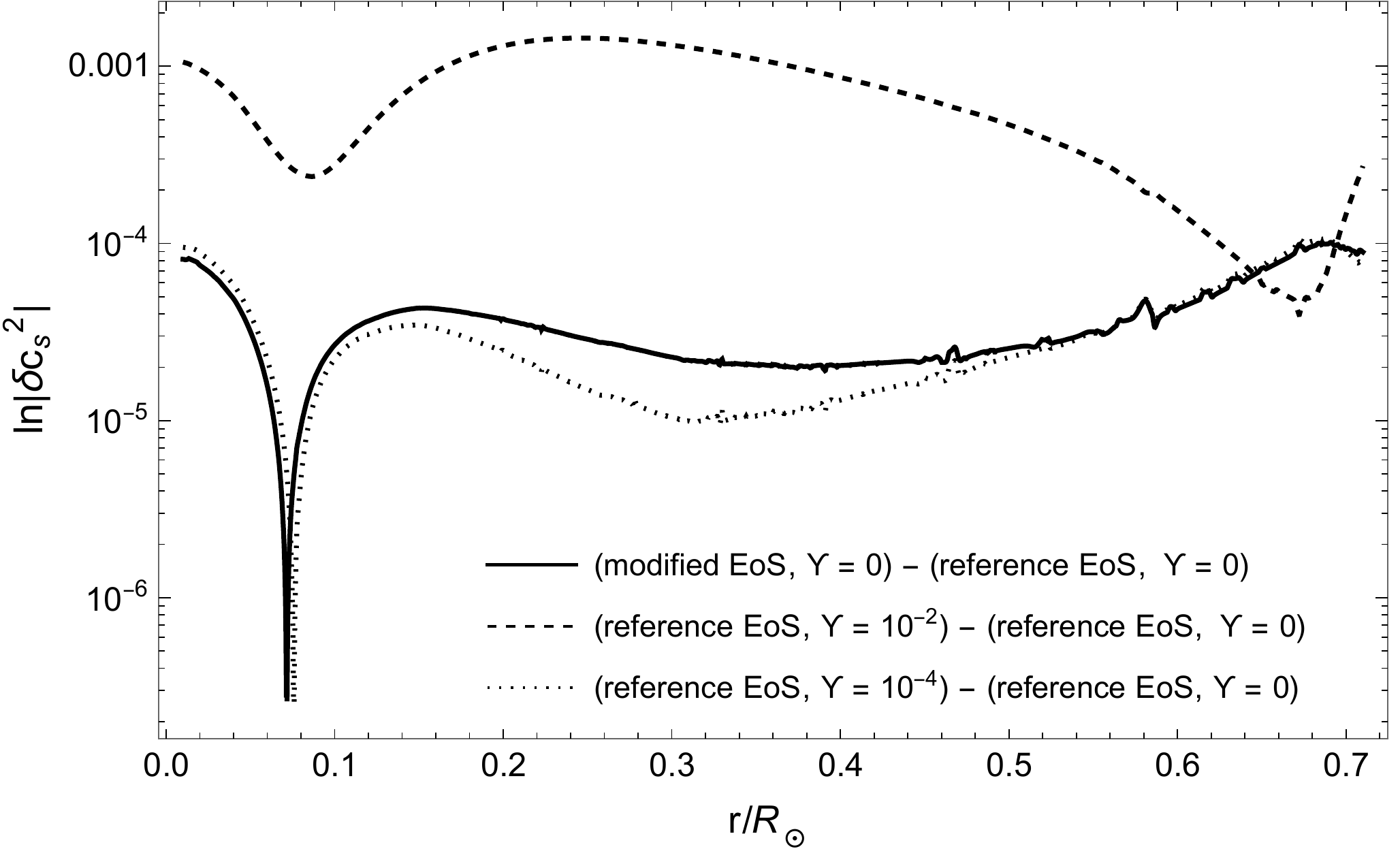}
	 \caption{ \label{fig:EoS} Comparison between the effect of a different EoS choice at standard gravity, and that of the fifth force at fixed EoS on the sound-speed profile. The variation $\delta \csq$ is in the sense (reference model) - (modified model). Our reference model uses the OPAL EoS, whereas the alternative EoS choice at standard gravity is the so--called Free EoS (see Section~\ref{sec:systematics}). The models with fifth force in the figure are computed on top of our reference model with OPAL EoS. It is seen that for a fifth-force coupling $\YMG = 10^{-2}$, the effect of a different EoS on the sound-speed profile remains significantly smaller than that of the fifth force, while for $\YMG = 10^{-4}$ the two effects start becoming indistinguishable. We remind that the fifth-force effect on the sound-speed profile within the radiative zone peaks around $r \simeq 0.25 R_{\odot}$ (see also Figures~\ref{fig:model_differences}, \ref{fig:delta_cs_components} and \ref{fig:delta-cs-real-syst}). Although the choice of high-level EoS does not constitute a systematic uncertainty in stellar modelling, the above result suggests that for sufficiently small values of $\YMG$, the choice of EoS matters. We emphasize that the good agreement between our reference EoS and the alternative one (black, continuous curve) highlights the consistency of our reference EoS in modelling the solar microphysics. }
\end{figure}

To test these effects, we produce a solar model at standard gravity and GS98 metallicity, but with the so-called Free EoS, instead of the OPAL which we used as our choice of reference. Figure~\ref{fig:EoS} shows a comparison between the choice of EoS at {\it standard gravity} and that of the fifth force at fixed EoS (OPAL) for the sound-speed profile. For $\YMG = 10^{-2}$ the fifth-force effect on the sound speed is significantly larger than that of the EoS deep in the radiative zone by a couple of orders of magnitude. However, for $\YMG = 10^{-4}$ the two effects become almost indistinguishable. Clearly, the choice of the EoS for sufficiently small values of $\YMG$ matters. This is, however, not an issue of concern. Typically, a choice for the EoS is made upfront in stellar modelling. As such, any EoS describing sufficiently well the relevant microphysics should not be considered to contribute to a systematic modelling uncertainty. Furthermore, the excellent agreement between our reference EoS (OPAL) and the alternative choice of Free EoS suggests the consistency of our choice of EoS, that is, our results are not expected to be biased by the choice of EoS. Finally, we should note that relativistic effects of electrons close to the solar centre could potentially become comparable to the fifth-force effect. However, the EoS's we use already implement such effects, and therefore there is no concern about their interference with the fifth force.

\subsection{The solar metallicity problem in light of the fifth-force effect} \label{sec:metallicity_problem}

Our analysis has been based on the GS98 choice for the solar metals mixture. Here, we ask how the solar metallicity problem might affect our results. The tension between solar models and helioseismic observations for models computed with the revised abundances, \citet{A09} (A09), compared to those computed with the ``old" solar metallicity inference of \citet{GS98} (GS98) is shown in Figure~\ref{fig:calibration}. As seen there, the predicted sound-speed profile, based on the metal mixture A09, is in tension with the profile predicted by helioseismic data, compared to calibrated models based on the GS98 metallicity mixture. If one assumes that the revised metallicity mixture (A09) is correct, the problem has been commonly attributed to mismatch in opacity modelling based on the GS98 mixture. 

 \begin{figure}
\hspace{-0.3cm} \includegraphics[width=93mm]{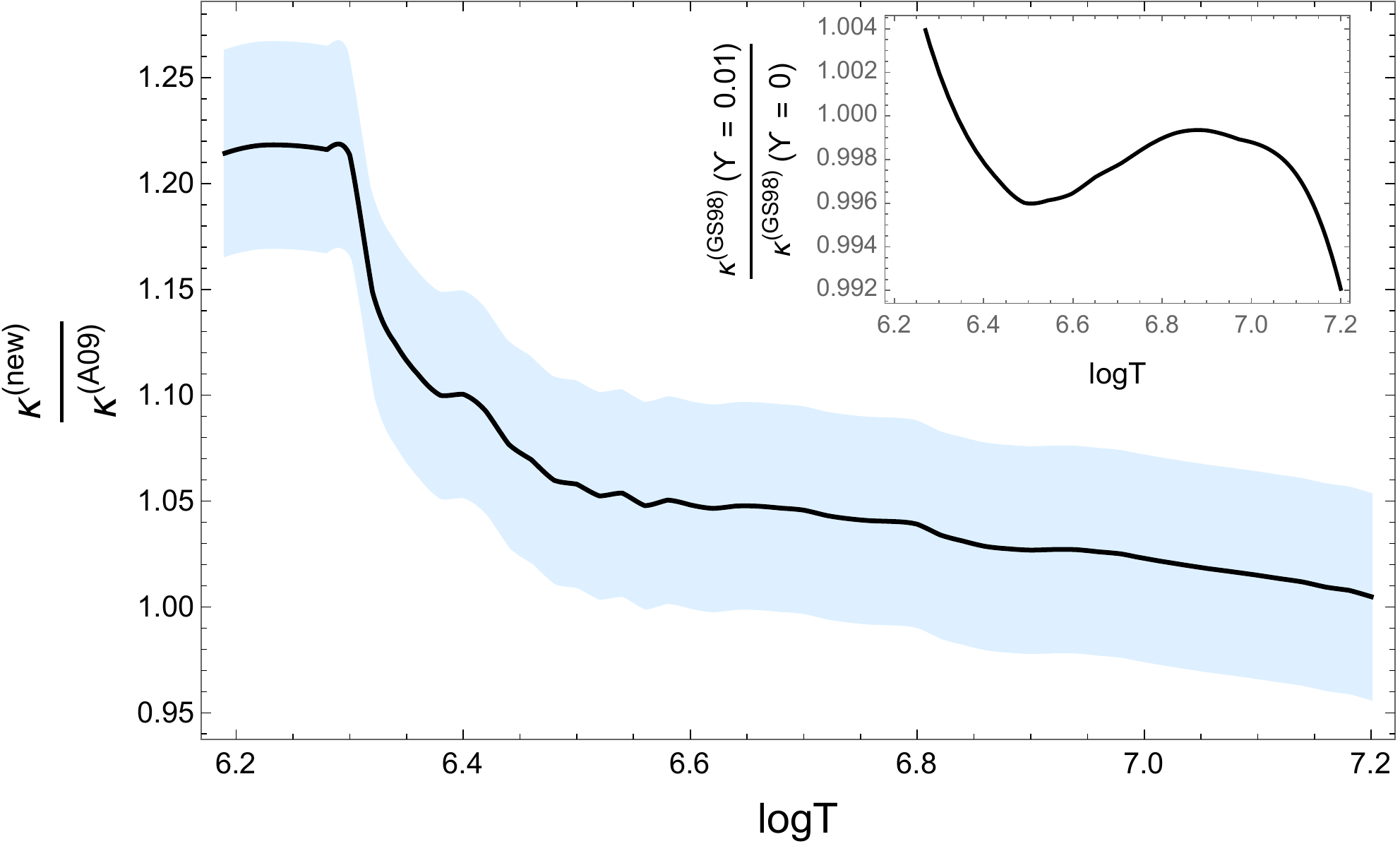}
 \caption{  \label{fig:Delta_kappa} An illustration of the opacity adjustment to reconcile the two metallicity mixtures, as explained in Section~\ref{sec:metallicity_problem}. Solar models computed with the revised mixture (A09) are in tension with helioseismic inferences, when compared to those computed with the ``old" mixture GS98 (see Figure~\ref{fig:calibration}, Table \ref{table:models} and text). A proposed approach to resolve the tension is to consider an appropriate opacity adjustment ($\kappa^{\text{(new)}}$) on top of the A09 profile ($\kappa^{\text{(A09)}}$), as shown in the above (black, solid curve). The shaded blue band shows the typical uncertainty of $\pm 5\%$, as discussed in Section~\ref{sec:systematics}. For comparison, the inset plot shows the induced change in the opacity profile under the fifth force computed with the GS98 metallicity mixture and unmodified opacity. Clearly, the induced opacity variation under the fifth force is significantly smaller and cannot interfere with the metallicity issue in any sense. The horizontal axis of log-temperature corresponds to a radial range between $\sim 0.0002 - 0.73 R_{\odot}$. The recent results of \cite{magg2022} suggest the solar metallicity issue is alleviated through the use of a revised metallicity mixture which is very close to the GS98 one.}
 \end{figure}

According to \citet{Christensen-Dalsgaard:2008idi, CD_Houdek:2009} the mismatch between the A09 and GS98 models can be bridged%
\footnote{As argued in those works, this proposal is mostly empirical and to large extent it lacks physical motivation.}
if we adjust the opacity profile of the A09 model $\kappa$ through a radius-dependent opacity correction $\Delta \kappa$ according to 
\be
\log \kappa^{\rm{new}} (\rho, T, X_i) =  \log \kappa (\rho, T, X_i) + \Delta \log \kappa, \label{eq:Delta kappa}
\ee
with the $\Delta \log \kappa$ given by
\be
\Delta \log \kappa \simeq \delta_{T} \log \kappa - \left( \frac{\partial \log \kappa^{\rm{GS98}}}{\partial \log \rho}\right)_{T,X} \delta_{T} \log \rho.
\ee
Here, $\delta_{T} \log \kappa \equiv \kappa^{\rm{GS98}}(\rho,T,X) - \kappa^{\rm{A09}}(\tilde \rho, T, \tilde X)$ is computed at constant temperature, and a similar relation holds for  $\delta_{T} \log \rho$. The result of this procedure is shown in Figure~\ref{fig:Delta_kappa}, where it is seen that the required change climbs up to $\sim 30 \%$ around the base of the convective zone. For comparison, it also shows the induced opacity under the fifth force computed with the old, GS98 mixture. The latter, is orders of magnitude smaller than the former, and therefore does not interfere with the proposed opacity adjustment. Certainly, the metallicity problem does not constitute a systematic effect similar to opacity or diffusion, and should be rather seen as a choice we make upfront; the fifth-force effect on interior profiles would manifest in a similar manner, had we chosen to work with the opacity-adjusted A09 mixture, instead of the GS98 one. 

Recently, \cite{magg2022} re-visited the solar metallicity problem. Their analysis suggests that the tension is reduced provided one adopts a newly proposed composition which is very close to the metal fractions of GS98. If this result is confirmed, it would imply that the solar metallicity problem is practically alleviated. For our analysis, these results support our choice of the GS98 metals fraction.

\begin{figure}
 \includegraphics[width=90mm]{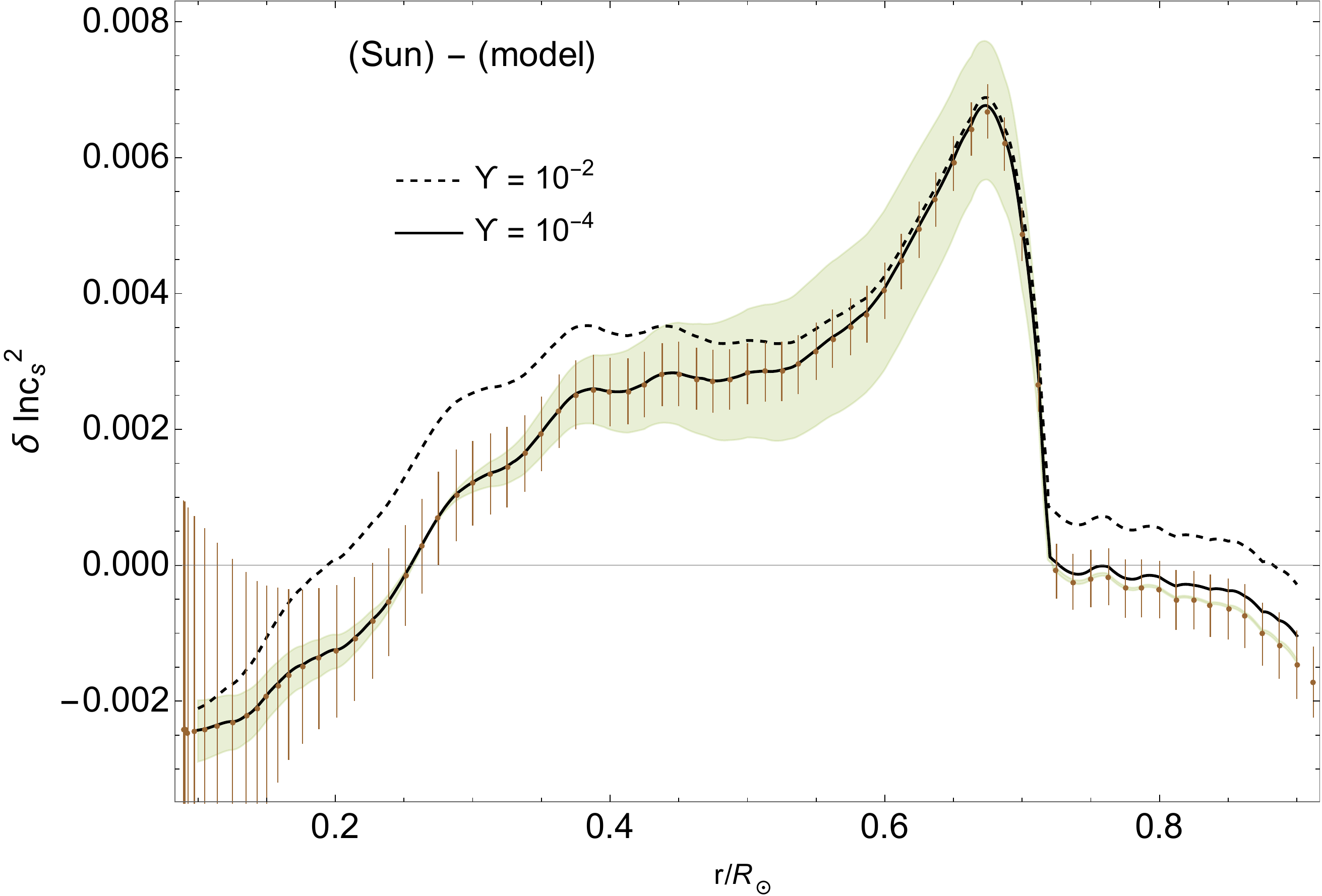}
	\caption{\label{fig:delta-cs-real-syst} Difference between the predicted (model) sound-speed profile and the profile derived from the inversion of helioseismic data (Sun), in the sense (Sun) - (model).  The brown vertical bars show the errors on the inverted helioseismic data points, according to the results of \cite{inversion_data}.  Errors have been multiplied by a factor of $10$. The shaded band corresponds to the maximum uncertainty from our modelling of opacity and diffusion, while the continuous and dashed curves to a modified model computed with a fifth-force coupling of $\YMG = 10^{-2}$ and $10^{-4}$ respectively. Fifth-force models are computed on top of our reference GS98 model at standard gravity (see Table \ref{table:models}). The fifth-force effect in the radiative zone is clearly seen for $\YMG = 10^{-2}$, as expected from the model differences of Figures~\ref{fig:model_differences} and \ref{fig:delta_cs_components}. Evidently, models computed with a fifth-force coupling $\YMG \lesssim 10^{-4}$ start being in agreement with helioseismic predictions for the sound-speed profile.}
 \end{figure}

\section{Constraints on the fifth-force coupling}  \label{sec:constraint}

Based on our previous equilibrium modelling, we now discuss how the two key solar regions, namely the radiative and convective zone, can be used to constrain generic gravity models. As we demonstrated earlier, each of these regions is impacted by input physics uncertainties in fundamentally different ways. This aspect is crucial, since it offers for complementarity of tests. Although our main interest is in the general scalar-field extensions of GR, our discussion easily extends to different theory setups, and highlight the significance of each solar region for precision tests of gravity.

\subsection{Radiative zone}
In this region, uncertainties from input physics tend to have a significant impact on interior profiles, most notably the sound speed as also seen from Figure \ref{fig:model_differences}. However, as we showed, the existence of a narrow region where our relatively simple estimates of modelling systematics become minimal offers an ideal place to test gravity (see Figure \ref{fig:delta_cs_components}). What is more, helioseismic inversions yield highly accurate results in this region, which implies that helioseismically inferred interior profiles show similar level of accuracy. 

Focusing on our particular theory setup parametrised by the fifth-force coupling $\YMG$, we denote with $c_{\text{s}(\odot) }^{2}$ the solar sound-speed profile inferred from inversions of helioseismic data (see also Figure \ref{fig:delta-cs-real-syst}), and $c_{\text{s}(\text{model})}^{2} \equiv  c_{\text{s}(\text{m})}^{2}(\YMG)$ the profile predicted by the GS98 model with $\YMG \neq 0$ at unmodified opacity/diffusion. To compare the model with the helioseismically inferred sound speed, we will employ a simple likelihood functional defined as
\begin{equation}
\mathcal{L}(\YMG) \propto e^{-\chi^2(\YMG)/2}, \label{eq:likelihood}
\end{equation}
with the $\chi^2$ constructed with the fractional change of the sound speed as
\begin{equation}
\chi^2 (\YMG) = \sum_{r_{i} = r_1}^{r_2} \frac{1}{\sigma_{i}^2} \left( \frac{ c_{\text{s}(\odot) }^{2}(r_{i}) }{ c_{\text{s}(\text{m})}^{2}(r_{i}; \YMG) }- 1  \right)^2. \label{eq:chi2-Y}
\end{equation}
$r_{i}$ denotes the radial points at which the $\chi^2$ is evaluated, between the radial boundaries $r_1, r_2$. We choose not to interpolate the discreet inverted data for the sound-speed profile ($c_{\text{s}(\odot) }^{2}$), as derived from \cite{inversion_data}.
Indeed, given the strong correlation between solutions at neighbouring points \citep[e.g.][]{Howe_Thompson:1996} such interpolation would have little meaning; we neglect this correlation in the present preliminary analysis, but it should be taken into account in future more detailed work.
We choose for the limits $r_{1},r_{2}$ approximately the solar region in the radiative zone where our systematics of opacity and diffusion become minimal (see Figures~\ref{fig:delta_cs_components} and \ref{fig:delta-cs-real-syst}) $(r_{1}, r_{2}) = (0.2, 0.3)$. In principle, any systematic uncertainties should be included in the statistical analysis and marginalised over in the end. In the particular solar region we are interested in though, the uncertainties from simple estimates of errors in the input physics are typically smaller than the helioseismic errors (see Figure \ref{fig:model_differences}). For this solar region, we have eight data points for the inverted sound-speed profile available to us. Evaluating (\ref{eq:likelihood}), we find that its maximum lies at $\YMG \simeq - 8 \cdot 10^{-4}$. Integrating it as usual to find the $2\sigma$ confidence interval and assuming a flat prior on $\YMG$, we find  
\begin{equation}
 - 10^{-3} \lesssim \YMG \lesssim  5 \cdot 10^{-4} . \label{eq:Y-bound}
\end{equation} 

This constraint on the fifth-force coupling strength for the general scalar-field theories improves previous bounds from stellar astrophysics by about three orders of magnitude. 
\\

It is now important to discuss and challenge our assumptions that led to the above result. 
 \begin{itemize}
 \item We have assumed the use of the OP opacity tabulation, which {along with the OPAL opacity table is one of the most advanced opacity tabulation to date}, and modelled the opacity error to be constant throughout the star ($= \pm5 \%$). This is a popular choice in the literature and appears to be a somewhat conservative strategy.  However, as we discussed in Section \ref{sec:systematics} (see also Figure \ref{fig:deltacs-opacity}), one cannot exclude the scenario where the opacity uncertainty acquires a radial dependence which peaks around $r \simeq 0.3 R_{\odot}$, that is, in the solar region within the radiative zone where the fifth-force effect peaks. Such an effect would interfere with the fifth-force effect in the solar radiative zone. We would like to entertain this scenario here. In order to understand the consequences of it, we assume the existence of a localised opacity uncertainty between $0.2 - 0.3 R_{\odot}$, which in turn induces a radially-dependent and localised change of the sound speed according to 
 \begin{equation}
 \frac{\delta c_{s}^2}{c_{s}^2} = \pm 10^{-3} \cdot e^{ - \frac{(r/R_{\odot} - 0.25)^2}{(0.0005)^2} }, \label{eq:Gaussian-delta-kappa}
 \end{equation}
 with the $\sigma = 5 \cdot 10^{-4}$ chosen so that the uncertainty fades off sufficiently fast close to the edges at $0.2$ and $0.3 R_{\odot}$ respectively. The chosen amplitude is about an order of magnitude smaller than the average (constant) uncertainty of $5\%$ we considered earlier, since it is already large enough to have a significant effect for our purpose. Notice that the $+$ and $-$ signs correspond to the case where the localised opacity uncertainty induces a positive (negative) effect on the sound speed. This is indeed a more general case that the one shown in Figure \ref{fig:deltacs-opacity}. We proceed repeating our previous likelihood analysis by adding on top of our theoretically computed sound speed the correction (\ref{eq:Gaussian-delta-kappa}). We find that for the ``+" and ``-" case in (\ref{eq:Gaussian-delta-kappa}) the following constraints at $2\sigma$; $1.8 \cdot 10^{-3} \lesssim \YMG \lesssim  3.8 \cdot 10^{-3}$, and $-4 \cdot 10^{-3} \lesssim \YMG \lesssim  -2\cdot 10^{-3}$ respectively. Both constraints are in tension with Newtonian gravity ($\YMG = 0$). This result is not surprising. A localised decrease (increase) of the sound speed, induced by a localised opacity uncertainty, requires a sufficiently positive (negative) fifth-force coupling $\YMG$. Indeed, $\YMG$ will tend to increase (decrease) the sound speed, compensating for the effect of the opacity change. In the presence of such a localised opacity uncertainty one would need to consider another solar region like the convection zone. We discuss this further below. From a statistical perspective, the opacity issue could be tackled through the exploration of the full opacity theory space along with that of the fifth force under a Monte-Carlo analysis. We comment further on this in Section \ref{sec:summary}. 
 \\
\item In this work, we have been interested in the solar equilibrium structure and the associated solar modelling systematics in the presence of the fifth force. In the absence of a proper helioseismic analysis in modified gravity, a key assumption we made is that the inverted solar sound-speed data we used as our ``observations" are computed at standard gravity. Indeed, helioseismic inversions infer the solar sound-speed profile from observed frequencies; however, being perturbative in nature they rely on the use of a background theoretical solar model. A known result in the literature is that the dependence of helioseismically inferred profiles have a weak dependence on the background standard solar model used \citep[see e.g.][]{Basu2000}. We expect that this result would carry over to some extent to the case of the particular gravity model studied here, given the smallness of the fifth-force coupling. However, only a fully consistent helioseismic analysis would provide a robust confirmation of our results, towards self-consistent constraints. Such an analysis is certainly necessary, and it will require the computation of solar frequencies and the extension of helioseismic inversions in the presence of the fifth force. 
 \end{itemize}
 

\subsection{Convective zone}
The solar convection zone offers yet another promising region to test deviation from Newtonian gravity. A key point here is that the scaling of systematics from input physics uncertainties is radically different compared to the radiative zone (see Figure \ref{fig:model_differences}). As it was explained in detail in Section \ref{sec:theory} using analytic relations, the reason behind this result is that the sound-speed profile in the convective zone is rather insensitive to the adiabatic structure of models with modified input physics. Therefore, the impact of variations under different input physics on the sound speed remains sufficiently small compared to the case of the radiative zone. 

For our particular gravity model, the fifth force leaves a sizable signature in this region, as seen in Figures \ref{fig:model_differences} and \ref{fig:delta-cs-real-syst}. A drawback of testing gravity within the solar convective zone is that helioseismic inversions do not work as well as they do within the radiative zone due to contamination from surface effects. This, however, can be circumvented with updated helioseismic inversion analyses based on observed solar frequencies of increased precision. Such an effort would promote the solar convective zone to a powerful region for tests of alternative gravity theories. Nevertheless, we find it a useful exercise to derive a constraint on $\YMG$ in the convective zone with our current helioseismic data which were derived assuming standard gravity. In this direction, we focus on the region $0.71 - 0.85 R_{\odot}$ in order to minimise potential biases from surface effects. Repeating our previous analysis, we find a preference for stronger gravity and $- 10^{-3} \lesssim \YMG \lesssim  10^{-5}$ at $2\sigma$. This constraint should not be taken seriously for the reasons explained earlier. A separate helioseismic analysis is needed in order to investigate the effect of the fifth force in the convective zone. However, the fact that the latter constraint is close to the one we found in the radiative zone provides some indication for the consistency of the analysis.

\section{Summary and future work} \label{sec:summary}
We employed the Sun as our laboratory to test general extensions of General Relativity. Although we focused on the case of scalar-field extensions of GR aiming to explain dark energy, our analysis easily applies to a broader set of gravity theories. By means of solar evolution simulations aided by analytical results, we presented a quantitative description of how the fifth force affects the equilibrium solar structure of calibrated solar models, and explained its interplay with the delicate solar microphysics such as opacity, diffusion, equation of state and metallicity (see e.g. Figure~\ref{fig:model_differences}). 

Despite the rather subtle competition between the fifth force effect and the solar modelling systematics,   we showed that the fifth force predicted by these general dark-energy theories leaves a sharp imprint on the solar sound-speed profile, with the peak of the effect located in a region where simple modelling uncertainties associated with opacity and diffusion tend to be insignificant. As we explained in Section~\ref{sec:systematics} and showed in Figure~\ref{fig:delta_cs_components}, the underlying mechanism for the latter result is due to the non-trivial scaling of temperature and mean molecular weight in the solar radiative zone under the influence of the fifth force, providing a distinct phenomenological imprint on the sound speed. As illustrated in Figure~\ref{fig:model_differences}, the sound speed in the solar convection zone is weakly affected by the general uncertainties in standard solar modelling, while the fifth force has a direct and significant effect. This offers an interesting prospect for future robust tests of fifth forces stemming from extensions of the standard paradigm of gravity and particles. Current inversions for the solar sound speed in this region suffer from some uncertainty in this region owing to the effects of the inadequate modelling of the near-surface layers \citep[e.g.,][]{JCD:1996}, but this may be alleviated through the use of more extensive data or improved analysis techniques.

A comparison with a helioseismically inferred sound-speed profile allowed us to derive a bound on the fifth-force coupling $\YMG$ which improves by approximately 3 orders of magnitude previous constraints from stellar physics (see Section \ref{sec:constraint}). This result relied upon the use of a helioseismically inferred sound-speed profile assuming standard gravity. 
A full analysis of the fifth-force effects will involve consistent calculations of solar oscillation frequencies in the presence of the fifth force, followed by an inverse analysis using the resulting frequencies.
In addition, to take other uncertainties of solar modelling consistently into account a Monte-Carlo analysis should be carried out including these uncertainties.
Such efforts will be reported in a future paper.

Our results highlight the power of the Sun as a laboratory for fundamental physics. In future, it would be very interesting to perform a helioseismic inversion analysis in the presence of the fifth force, extending the current machinery to the case of the fifth force. In turn, this will allow us to infer the sound-speed profile from observations in a fully consistent manner, and confirm the results presented in this work from a rather bottom-up helioseismic approach. Finally, it would be interesting to understand the impact of the fifth force on solar neutrinos and its confrontation with observations from earth-based detectors.

\begin{acknowledgements}
Ippocratis D. Saltas is supported by the Czech Grant Agency (GACR), under the grant number 21-16583M. Funding for the Stellar Astrophysics Centre is provided by The Danish National Research Foundation (Grant DNRF106).
It is a pleasure to acknowledge enlightening discussions with Earl Bellinger, Rasmus Handberg, Steen Hannestad, Stefan Ilic, Ilidio Lopes, Lorenzo Pizzuti. 

\end{acknowledgements}

\clearpage
\bibliographystyle{aa} 
\bibliography{master_jcd_v7x}

\end{document}